\documentclass{optica-article}

\journal{opticajournal} 

\articletype{Research Article}

\usepackage{lineno}
\usepackage{amsmath,amsfonts}
\usepackage{algorithmic}
\usepackage{algorithm}
\usepackage{array}
\usepackage{subfigure}

\usepackage{textcomp}
\usepackage{stfloats}
\usepackage{url}
\usepackage{verbatim}
\usepackage{graphicx}
\usepackage{cite}

\newcommand{\Fig}[1]{Fig.\,{\ref{#1}}}

\begin{document}

\title{Temporal Paraxial Optics under Adiabatic Modulations}

\author{Antonio Alex-Amor,\authormark{1,*} Carlos Molero\authormark{2}}

\address{\authormark{1}Department of Electronic and Communication Technology, RFCAS Research Group,  Universidad Autónoma de Madrid, 28049 Madrid, Spain\\
\authormark{2} Department of Electronic and Electromagnetism, Faculty of Physics, University of Seville, 41012, Seville, Spain
}

\email{\authormark{*}antonio.alex@uam.es} 


\begin{abstract*} 
This paper presents a temporal paraxial formulation for the propagation of ultrashort optical pulses in time-modulated media with slowly varying refractive index. By deriving the paraxial wave equation directly in the time domain from the Helmholtz equation under an adiabatic approximation, the model remains analytically tractable while extending paraxial optics beyond time-invariant backgrounds commonly treated by frequency-domain expansions. The resulting equation preserves a Schrödinger-like structure in the presence of explicit temporal modulation and admits closed-form solutions for ultrashort Gaussian pulses. The framework supports a Green’s-function description and an operator-based Hamiltonian formalism, from which an ABCD matrix representation for temporal propagation in time-varying media is obtained. The results demonstrate that temporal modulation provides a dynamic means to control ultrashort pulse dynamics, enabling tailored evolution of pulse characteristics such as temporal width and chirp, with potential applications in ultrafast pulse shaping and a direct connection to temporal wave-packet dynamics.  

\end{abstract*}

\section{Introduction}
Paraxial optics constitutes an approximation within Gaussian optics and ray tracing whose central assumption is the small-angle propagation of light beams with respect to the optical axis~\cite{GoodmanBook}. Under this approximation, the analysis of complex optical systems is greatly simplified, as the relationships between input and output variables become linear, enabling matrix-based descriptions and simpler geometrical rules~\cite{Mouroulis1996, wolf2004geometric}. The connection to wave optics is established by associating individual rays with local wavefronts. Within the paraxial regime, the spatial evolution of optical fields is governed by the well-known paraxial wave equation~\cite{Melvin1975}, which is obtained by applying the slowly varying envelope (or wave-function) Ansatz to the Helmholtz equation~\cite{Siegman1986}. The resulting equation takes a Schrödinger-like form~\cite{Moya2025, Torres2011, HausModeLocking1975}, in which the longitudinal derivative of the field with respect to the propagation coordinate  is proportional to the second-order derivatives with respect to one or both transverse coordinates~\cite{Marte1997}.

A straightforward duality can be established with a temporal counterpart, in which the role of the transverse coordinate $y$ in the paraxial equation is replaced by time $t$~\cite{Agrawal2025, Kolner1994, Akhmanov1969}. In this framework, the equation describes the dispersion of a short temporal pulse as it propagates along $z$, corresponding to the case of narrow-band dispersion. This duality arises from the mathematical equivalence between the spatial and temporal domains and has proven useful in extending optical concepts into the temporal regime. Temporal optics encompasses phenomena such as time lenses and time prisms~\cite{Howe2006}, whose temporal function is analogous to the action of a spatial lens or prism. Experimental implementations of time-lens architectures have been demonstrated~\cite{Godil1993, Azana2004, Shawn2009, Salem2013}, revealing parallels between spatial and temporal systems for concepts such as focal distance and magnification~\cite{Okawichi2009}. Additional analogies and applications include group-delay dispersion circuits~\cite{Sharma1994}, temporal solitons~\cite{AgrawalMDPI24}, pulse amplification and compression~\cite{STRICKLAND1985, Kolner1988}, and temporal lensing and chirping effects~\cite{EnghetaTemporalLensing2025}.

In this context, photonic and optical phenomena arising in media with a time-varying refractive index have recently attracted significant attention within the RF, optics, and photonics communities~\cite{Caloz2020_2, Ptitcyn2023, Pacheco2022}. Although the foundational ideas date back to the mid-20th century~\cite{Morgenthaler1958, Felsen1970, Fante1971}, recent progress in experimental capabilities, together with advances in theoretical frameworks, has driven a renewed surge of interest in this area. In particular, contributions from eminent researchers such as Prof. Nader Engheta, Prof. Andrea Alù, Prof. John Pendry, and Prof. Christophe Caloz, among others, have enabled new approaches to engineering wave–matter interactions in the time domain, leading to applications including frequency conversion, nonreciprocity, and amplification~\cite{engheta2023four, Galiffi2022, Caloz2020_2}.

Moreover, the exploration of dual scenarios obtained by interchanging the roles of space and time has emerged as a powerful framework for deeper conceptual understanding, as highlighted in recent works on temporal metamaterials by Profs. Pachecho-Peña and Engheta \cite{EffectiveMediumEngheta2020}. This approach has enabled the development of temporal counterparts to well-known spatial phenomena, including the temporal Brewster angle~\cite{Pacheco2021}, temporal Snell’s law~\cite{Mendoca2002}, temporal impedance matching~\cite{PachecoEngheta20, AluImpedance2019}, as well as time refraction, reflections and diffraction at abrupt temporal interfaces \cite{Zurita2009, Tirole2023SlitTime, Alex2023TV}. Within this same paradigm, photonic time crystals~\cite{Sharabi_2022, Galiffi2022, BoltassevaOpinion24}, inspired by their spatial analogues, have attracted increasing attention, as a refractive index periodically modulated in time gives rise to band structures in the frequency domain and the formation of momentum bandgaps. Under suitable conditions, such systems can exhibit physical unconventional phenomena that opens new opportunities for wave manipulation in time-varying media~\cite{Xiong2025}.

In the context of optics, to the authors’ knowledge, most existing derivations of the temporal paraxial equation consider time-invariant background media, which may be dispersive but do not vary explicitly in time~\cite{Weisser2026, Agrawal2025}. In these works, the propagation constant is expanded to second order around a central frequency under the paraxial approximation, yielding a frequency-domain equation. The temporal version can then be obtained via inverse Fourier transform, resulting in an analytically tractable equation. Importantly, the frequency-dependent refractive index, $n(\omega)$, does not imply any intrinsic time dependence of the medium: the time-domain pulse may evolve, experiencing chirp or broadening due to dispersion, but the medium itself remains time-invariant. While this approach is adequate for slightly-dispersive, stationary media, it becomes less practical when the medium exhibits strong dispersion or explicit temporal modulation. 

Conversely, formulating the problem directly in the time domain allows a more general treatment, extending naturally to nonlinear systems and providing direct physical insight~\cite{Bennet1978}. In particular, it enables the study of background media whose refractive index varies explicitly in time, $n = n(t) = \sqrt{\varepsilon(t) \mu(t)}$~\cite{Mirmoosa2022, Memarian2024}, which can induce phenomena beyond standard dispersive chirping.

A particularly relevant scenario arises when considering ultrashort pulses propagating in time-varying media. Ultrashort optical pulses have become an indispensable tool in modern photonics, enabling access to physical processes occurring on picosecond and femtosecond timescales and supporting applications such as ultrafast spectroscopy, coherent control~\cite{Weiner2011}, high-precision material processing~\cite{Gattass2008}, optical metrology, and ultrafast signal processing~\cite{Capmany2007}. In this context, the temporal modulation of the material can often be regarded as locally quasi-static, since the  duration of the ultrashort pulse is typically orders of magnitude shorter than the characteristic time scale of the modulation, allowing the pulse to retain most of its geometric and spectral properties during propagation. This situation corresponds to the so-called \emph{adiabatic regime}, a concept inspired by the quantum adiabatic theorem~\cite{Born1930, kato1950}, in which the medium evolves slowly compared to the pulse dynamics. Within this regime, the paraxial approximation can be extended over finite temporal and spatial intervals while preserving the validity of the analysis. 

This paper explores the mathematical and physical implications of a paraxial optical system evolving in a slowly-varying \emph{time-modulated} medium, which is assumed to be non-dispersive, i.e., the medium response is instantaneous. Thus,  pulse evolution is affected by the explicit time dependence rather than by dispersive reshaping arising from material memory. 

\begin{figure}[!t]
    \centering      \subfigure{\includegraphics[width=0.98\columnwidth]{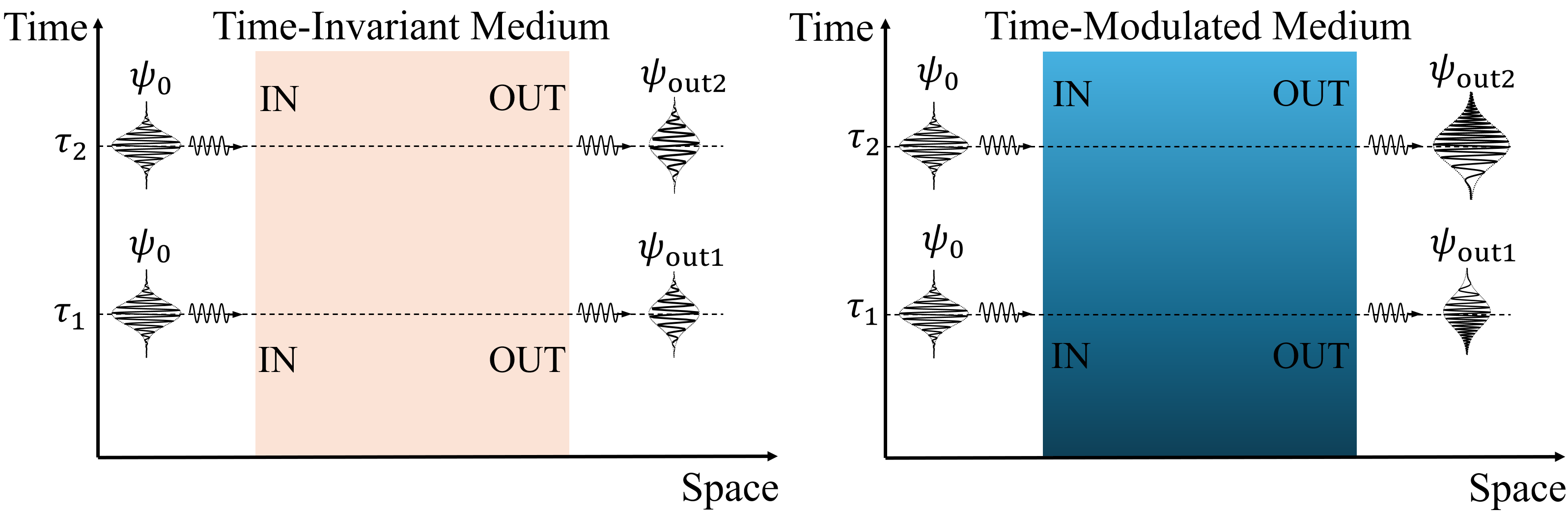}}
    \caption{Propagation and diffraction of temporal pulse across a time-invariant (left) and a time-modulated (right) medium. The time-modulated scenario offers a dynamical control on the pulse characteristics (duration, chirp, etc.) as the medium parameters change in time. On the other hand, the time-invariant medium shows the same response independently of the chosen time instant. }
    \label{fig1}
\end{figure}

The corresponding temporal paraxial wave equation is then derived by applying the slowly-varying envelope Ansatz to the Helmholtz equation in the time domain. Under the adiabatic condition, the mathematical complexity of the equation is reduced, yielding a generalized temporal wave equation that preserves the Schrödinger-like structure. This formulation still admits solutions in the form of ultrashort Gaussian pulses and allows the redefinition of the associated Green's function to describe propagation under temporal sources. This new formulation is applied to describe the evolution of ultra-short Gaussian pulses in time-variant media. \Fig{fig1} highlights the differences between pulse dynamics in time-invariant and time-variant media and the opportunities that the latter offer. The time-variant case introduces additional degrees of freedom in the system, allowing for a dynamical control of the pulse characteristics such as the duration or the chirp. This new scenario can be identified with the evolution of a particular case of space-time wave packets \cite{Yessenov:22}. 

Furthermore, the equivalence of the equation to a Schrödinger system enables a formulation in terms of mathematical operators~\cite{Moya-Cessa_2025} and the definition of an analogous Hamiltonian for the system~\cite{LIU2000}. Interpreting the system via the Hamiltonian is advantageous: the original adiabatic theorem was formulated for Hamiltonian systems~\cite{Duan2020}, and Hamilton's equations provide a direct route to determine system dynamics and analytical relationships between time and frequency. This operator-based approach also facilitates an ABCD matrix formalism, which is particularly useful for connecting (linearly) input and output variables in a compact and physically intuitive way.  

\section{Temporal Paraxial Wave Equation}

In the following, we work with linear, isotropic, and non-dispersive media governed by the constitutive relations $\mathbf{D} = \varepsilon \mathbf{E}$ and $\mathbf{B} = \mu \mathbf{H}$. From a macroscopic perspective, temporally-modulated media can be described by means of the time-varying permittivity $\varepsilon(t) = \varepsilon_{0}\varepsilon_{\text{r}}(t)$ and permeability $\mu(t) = \mu_{0} \mu_{\text{r}}(t)$ functions. For simplicity, we may omit the time dependence of $\varepsilon$ and $\mu$ in the derivations. Moreover, we assume that the fields are linearly-polarized according to $\mathbf{E} = \hat{\mathbf{x}}E_x$, $\mathbf{H} = \hat{\mathbf{y}}H_y$ and follow the convention $\mathrm{e}^{-i\omega_0 t} \mathrm{e}^{+ik_0 z}$. 

By operating from Maxwell's equations in the present scenario, the scalar wave equation that describes the evolution of the electric field reads
\begin{equation} \label{scalar_waveequation}
    \frac{\partial^2 E_x}{\partial z^2}  = \frac{\partial}{\partial t}  \left[ \mu\, \frac{\partial \left(\varepsilon E_x \right)}{\partial t}   \right]\, .
\end{equation}

In analogy to Gaussian-beam propagation in the spatial case, we may assume that a time-modulated Gaussian beam  propagating in space, $E_x(t,z)$, is described by a field envelope function $\psi(t,z)$ modulated by a plane-wave carrier:
\begin{equation} \label{Ex_time}
    E_x(t,z) = \psi(t,z)\, \mathrm{e}^{-i\omega_0 t} \mathrm{e}^{+ik_0 z}\, .
\end{equation}
The free-space wavenumber $k_0$ and carrier frequency $\omega_0$ are related through $c = \omega_0 / k_0$, $c$ being the speed of light in vacuum. The plane-wave carrier is chosen as an invariant reference, while the effects of the medium and transverse structure are fully captured by the evolution of the envelope.

The insertion of eq. \eqref{Ex_time} into the scalar wave equation \eqref{scalar_waveequation} leads to
\begin{multline} \label{wave_equation_time1}
    \frac{\partial^2 \psi}{\partial z^2} + i2k_0 \frac{\partial \psi}{\partial z} - k_0^2 \psi = 
    \mu \varepsilon \left[\frac{\partial^2 \psi}{\partial t^2} - i2\omega_0 \frac{\partial \psi}{\partial t} - \omega_0^2 \psi \right] \\ 
    + 
    \left[\frac{\partial \psi}{\partial t} - i\omega_0 \psi \right] \left[ \varepsilon \frac{\partial \mu}{\partial t} + 2\mu \frac{\partial \varepsilon}{\partial t} \right] 
     + 
    \left[\mu \frac{\partial^2 \varepsilon}{\partial t^2}+ \frac{\partial \mu}{\partial t} \frac{\partial \varepsilon}{\partial t} \right] \psi\, .
\end{multline}
The details of the derivation are in the Supplementary Material. 

In principle, the function $\psi(t,z)$ can be of arbitrary value. However, we are interested in replicating the \emph{paraxial conditions} seen in conventional spatial wave optics. To do so, we must consider that:
\begin{enumerate}
    \item  $\psi(t,z)$ is a slowly-varying function along the propagation direction $z$. Under this condition, $|\partial^2 \psi / \partial z^2| \ll |k_0\cdot  \partial \psi / \partial z|$, thus the second spatial derivative term can be neglected.  
    \item The time-modulated permittivity and permeability functions vary slowly with time; namely, their variation is slow compared to the carrier frequency $\omega_0$. Under these conditions, we may neglect the time derivatives of $\varepsilon(t)$ and $\mu(t)$.  
\end{enumerate}

Under these assumptions, the notion of \emph{paraxiality} can be naturally extended to the temporal domain. In particular, requiring that the permittivity and permeability be slowly modulated in time implies that the modulation frequency $\omega_m$ is much smaller than the carrier frequency $\omega_0$, i.e., $\omega_m \ll \omega_0$. Importantly, this condition is readily achievable in practice and is widely exploited not only in optics and photonics, but also in time-varying systems such as time-modulated circuits, waveguides, and metamaterials, where the modulation frequency is typically orders of magnitude lower than the operating (carrier) frequency \cite{Melcon2019, Zhang2018TimeModulated}.

Thus, in the \emph{temporal paraxial regime},  the temporal wave equation \eqref{wave_equation_time1} simplifies after some algebraic manipulations to
\begin{equation} \label{wave_equation_time2}
    \frac{\partial \psi}{\partial z} \approx - \frac{i}{2k_0 v^2}\frac{\partial^2 \psi}{\partial t^2} - \frac{c}{v^2} \frac{\partial \psi}{\partial t} - \frac{i\omega_0}{2c} \left(1 - \frac{c^2}{v^2} \right) \psi\, .
\end{equation}
The parameter $v = v(t) = [\varepsilon(t)\mu(t)]^{-\frac{1}{2}}$ stands for the propagation velocity in the time-modulated host medium. It is related to the time-varying refractive index through $v(t) = c/n(t)$. Notice that, for time-invariant free-space propagation conditions ($v(t) = c$), eq. \eqref{wave_equation_time2} reduces to $\partial \psi / \partial z \approx - i / ( 2 \omega_{0} c) \times  \partial^2 \psi / \partial t^2 - 1/c \times  \partial \psi / \partial t$, as that in \cite{Agrawal2025}.

Instead of working with the laboratory time $t$,  it is more convenient to use the retarded/local time $\tau = t - z c/v^2$. In this frame, $E_x(\tau, z) = \psi(\tau, z) \mathrm{e}^{- i\omega_0\tau } \mathrm{e}^{+ ik_0z} 
\mathrm{e}^{- i z \omega_0c/v^2}$. From a physical perspective, this change of variable implies co-moving at the same speed than the pulse. Mathematically, it allows us to cancel the second right-hand term in eq. \eqref{wave_equation_time2} (after neglecting the time derivative of $v$) and further reduce the temporal paraxial wave equation to
\begin{equation} \label{wave_equation_time3}
    \frac{\partial \psi(\tau, z)}{\partial z} \approx - ia(\tau) \frac{\partial^2 \psi(\tau,z)}{\partial \tau^2}  - ib(\tau) \psi(\tau, z)\, ,
\end{equation}
with time-dependent coefficients
\begin{equation} \label{a_b}
    a(\tau) = \frac{1}{2k_0 v^2(\tau)}, \quad b(\tau) = \frac{\omega_0}{2c} \left(1 - \frac{c^2}{v^2(\tau)} \right).
\end{equation}

Equation~\eqref{wave_equation_time3} is closely related to Schrödinger-type evolution equations that are widely employed in the theory of pulse propagation in dispersive media~\cite{AgrawalBook}. In the conventional treatment, such equations are derived under the assumption of a time-invariant background medium. The propagation constant $\beta(\omega)$ is expanded in a Taylor series around a carrier frequency $\omega_0$, and the resulting paraxial equation is formulated in the frequency domain. A temporal evolution equation is then obtained by applying an inverse Fourier transform~\cite{Kolner1994, Agrawal2025, AgrawalMDPI24}. Within this framework, temporal effects arise solely from material dispersion encoded in the frequency dependence of $\beta(\omega)$, while the medium parameters themselves remain stationary.

The approach adopted here differs from the previous paradigms by working directly in the time domain, instead of employing frequency-domain expansions tailored to time-invariant background media.
This allows explicit temporal modulations of the medium parameters to be incorporated from the outset. By introducing a retarded time coordinate, the dynamics are simplified while preserving the full time dependence of the effective wave velocity $v(\tau)$, which directly determines the coefficients $a(\tau)$ and $b(\tau)$ in eq.~\eqref{a_b}. As a result, temporal variations of the medium influence the pulse evolution through an explicit time
dependence of the governing evolution operator, even in the absence of dispersion.

\subsection{Connection to Conventional Spatial Paraxial Optics}

As will be shown, there exists a close formal correspondence between temporal and spatial propagation problems. As a consequence, many well-established concepts from spatial paraxial optics, such as diffraction, beam-waist evolution, and effective potentials, admit direct temporal analogues. This correspondence provides a powerful framework that greatly simplifies both the analysis and the physical interpretation of time-varying optical phenomena.

In conventional treatments of spatial paraxial optics, the surrounding medium is typically assumed to be homogeneous and time-invariant, with constant permittivity and permeability, i.e., $\varepsilon=\mathrm{const.}$ and $\mu=\mathrm{const}$ \cite{Pedrotti2007, Hecht2017}. However, a complete analogy with time-modulated optical systems emerges only when the spatial problem is generalized to include material inhomogeneities.

Accordingly, we consider a spatial paraxial configuration in which the medium is deliberately modulated along the transverse direction, with constitutive parameters of the form $\varepsilon=\varepsilon(y)$ and $\mu=\mu(y)$. Within this generalized framework, the transverse spatial coordinate $y$ plays a role formally equivalent to that of time $t$ in temporal optics. Under this mapping, spatial material modulations correspond directly to temporal modulations of the medium, enabling the direct transfer of intuition and analytical tools between spatially structured and time-varying optical systems.

It can be shown that the evolution of a spatially-modulated time-harmonic Gaussian beam of the form $E_x(y,z) = \psi(y,z) \mathrm{e}^{+ ik_0 z}$ ($\mathrm{e}^{- i\omega_0 t}$ is implicit and therefore omitted), which is slowly-varying along the propagation direction $z$, is described via the wave equation
\begin{equation} \label{wave_equation_space}
    \frac{\partial \psi(y,z)}{\partial z}  \approx +\frac{i}{2k_0} \frac{\partial^2 \psi(y,z)}{\partial y^2}  - \frac{i \omega_0}{2c} \left(1 - \frac{c^2}{v^2(y)} \right) \psi(y,z).
\end{equation}
The reader is referred to the Supplementary Material to see the details of the derivation. The spatial paraxial equation can then be reduced to $\partial \psi / \partial z = -ia \partial^2 \psi / \partial y^2 - ib(y) \psi$.

As observed, the temporal \eqref{wave_equation_time3} and spatial \eqref{wave_equation_space} paraxial wave equations share the same mathematical form. This implies that, in the paraxial regime discussed in the work, the temporal Gaussian beam  defined by eq. \eqref{Ex_time} is expected to propagate and diffract in a manner very similar to that of a well-known spatial Gaussian beam. Nonetheless, a close inspection of \eqref{wave_equation_time3} and \eqref{wave_equation_space} reveals some minor differences: 1) The difference in the sign of the first right-hand term in both equations simply comes from the phase convention of the carrier and the nature of the vector operators ($\nabla^2$ vs $\partial^2_t$). 2) Although the coefficient $b$ is dependent on the transverse variable in the spatial [$b = b(y)$] and temporal [$b = b(\tau)$] scenarios, the coefficient $a$ is not just in the spatial case. This difference lies on how the spatial and temporal derivatives show up in Maxwell's equations: spatial derivatives act exclusively on field components while temporal derivatives act on both material parameters and fields. Therefore, although spatial and temporal paraxial wave equations are formally analogous, the dependence of their coefficients encodes a fundamental physical distinction: transverse spatial modulations perturb the dispersion relation additively with respect to a fixed carrier, whereas temporal modulations continuously reshape the local dispersion relation $\omega-k$. Table 1 summarizes the mapping between quantities in both spatial paraxial and temporal paraxial optics.

\begin{table}[h!]
\centering
\begin{tabular}{c c c}
\hline
Spatial Paraxial Optics &  & Temporal Paraxial Optics \\
\hline
$y$ & $\leftrightarrow$ & $\tau$ \\
$z$ & $\leftrightarrow$ & $z$ \\
$\psi(y,z)$ & $\leftrightarrow$ & $\psi(\tau,z)$ \\
$v^2(y) = [\varepsilon(y) \mu(y)]^{-1}$  & $\leftrightarrow$  & $v^2(\tau) = [\varepsilon(\tau) \mu(\tau)]^{-1}$\\
$a = -\frac{1}{2k_0}$ & $\leftrightarrow$ & $a(\tau) = + \frac{1}{2k_0 v^2(\tau)}$ \\
$b(y) = + \frac{\omega_0}{2c} \left(1 - \frac{c^2}{v^2(y)} \right)$ & $\leftrightarrow$ & $b(\tau) = + \frac{\omega_0}{2c} \left(1 - \frac{c^2}{v^2(\tau)} \right)$\\
\hline
\end{tabular}
\caption{Mapping between spatial and temporal paraxial optics 
quantities.}
\end{table}

\subsection{Connection to Quantum Mechanics}

The temporal paraxial wave equation \eqref{wave_equation_time3} has the mathematical structure of a Schrödinger-like equation \cite{GriffithsBook}. This can be clearly seen by multiplying both terms of eq. \eqref{wave_equation_time3} by $+i$, with 
\begin{equation}\label{Schro}
    \hat{H} = -a(\tau) \frac{\partial^2 }{\partial \tau^2} - b(\tau)    \, 
\end{equation} 
representing the analogous Hamiltonian of the optical system \cite{Fuch2022, Agrawal2025}.  
In the temporal optics scenario, the evolution variable is $z$, while in quantum mechanics, it is time \cite{Longhi2009quantum}. Under this correspondence, the field envelope $\psi$ formally maps onto a quantum wavefunction, with the Hamiltonian operator $\hat{H}$ governing its evolution along the propagation direction $z$. The time-dependent coefficient $a(\tau)$ plays the role of an effective mass term, while $b(\tau)$ acts as an effective potential, both in time. In fact, a close inspection on the Hamiltonian structure in \eqref{Schro} reveals dimensions of wavenumber. Notice that, by multiplying both members of \eqref{Schro} by $\hbar$, the analogous Hamiltonian acquires dimensions of momentum, in accordance to quantum-physics operators when the evolution variable is spatial and not time \cite{Dias2025}. When $a$ and $b$ are independent of $\tau$, the analogous Hamiltonian reduces to that of a free particle in a constant potential, leading to dispersion-driven spreading of the optical waveform. Conversely, a nontrivial $\tau$-dependence of $a(\tau)$ or $b(\tau)$ gives rise to effective forces and confinement mechanisms analogous to those encountered in quantum systems with spatially-varying mass or external potentials.

Despite this formal equivalence, it is important to emphasize that the temporal paraxial equation describes a classical wave phenomenon. The Schrödinger-like structure arises from the paraxial and slowly-varying envelope approximations and does not imply quantum behavior of the optical field itself. Nevertheless, this analogy enables the transfer of intuition and analytical techniques from quantum mechanics, such as operator methods, eigenmode expansions, and adiabatic approximations, to the analysis of wave propagation and diffraction in structured temporal optical media.

\section{Adiabatic Green's Function Solution}
In this section, we derive a Green's Function solution to the temporal paraxial equation under temporal adiabatic modulations. Physically, the adiabatic regime is reserved for the propagation of \emph{ultra-short temporal pulses} in time-modulated media; namely, pulses whose duration $T_\mathrm{pulse}$ is much shorter than the modulation period $T_v$ of the medium parameters $\varepsilon(\tau)$ and $\mu(\tau)$, and thus of the propagation velocity $v(\tau)$. When $T_\mathrm{pulse} \ll T_v$, all the sections within the temporal pulse (center and tails) experience the time-modulated medium as quasi-static, perceiving only a slowly varying environment during its propagation. As a result, although the medium effectively changes its properties in time, the pulse \emph{locally} retains its overall shape while it propagates.

From a mathematical perspective, the temporal adiabatic regime, while more general than the commonly assumed time-invariant case, greatly simplifies both the calculations and the interpretation of the results. In particular, the derivatives of the time-dependent coefficients can be neglected: $\partial^n a(\tau) / \partial \tau^n = \partial^n b(\tau) / \partial \tau^n \approx 0 $, $n$ being the order of the derivative. This approximation allows us to factor out slowly-varying phase terms, treat the dispersive term independently, and apply analytical techniques such as Green's functions or ABCD matrix methods. Importantly, this adiabatic assumption ensures that frequency shearing, pulse broadening, and other modulation-induced effects can be treated in a controlled, perturbative manner, preserving the Gaussian character of the pulse and allowing for an elegant operator- or function-based description of its evolution.

In this regard, it should be noted that last right-hand term in eq. \eqref{wave_equation_time3}, $b(\tau)\psi(\tau, z)$, essentially represents a local phase term. This extra phase term can be removed from the equation with a substitution of the form 
\begin{equation}  \label{psi_chi}
    \psi(\tau, z) = \chi(\tau,z) \mathrm{e}^{-ib(\tau) z}\, .
\end{equation}
Under the adiabatic regime ($b$ only contributes locally and its time derivatives can be neglected), the temporal paraxial wave equation \eqref{wave_equation_time3} reduces to
\begin{equation} \label{wave_equation_time4}
    \frac{\partial \chi(\tau, z)}{\partial z} \approx - ia(\tau) \frac{\partial^2 \chi(\tau,z)}{\partial \tau^2} \, ,
\end{equation}
with $a$ given by eq. \eqref{a_b}. 

The extraction of $\chi(\tau, z)$ from the reduced paraxial equation \eqref{wave_equation_time4} allows us to build the complete field profile solution of $\psi(\tau, z)$ and $E_x(t,z)$ by using \eqref{psi_chi} and then \eqref{Ex_time}. By means of its Green's function $G_\chi(\tau, z)$,  $\chi(\tau,z)$ is computed at a retarded time $\tau$ and position $z$ when the source was located at time $\tau_0$ and position $z_0 = 0$:
\begin{equation} \label{convolution}
    \chi(\tau, z) = \int_{-\infty}^\infty G_\chi(\tau, z; \tau_0)\, \chi_0(\tau_0)\, d\tau_0\, ,
\end{equation}
where $\chi_0$ represents the initial condition or temporal source distribution.

It is shown in the Supplementary Material how the Green's function $G_\chi$ can be approximated in the temporal adiabatic regime by
\begin{equation} \label{Greens_function}
        G_\chi (\tau, z; \tau_0) \approx \frac{1}{\sqrt{- i4 \pi a(\tau) z}} 
\exp\Bigg[ - i \frac{(\tau-\tau_0)^2}{4 a(\tau) z} \Bigg]\, .
\end{equation}
This solution is exact for time-independent coefficients $a$, but approximate for temporal adiabatic scenarios. In the adiabatic regime, the variation of the coefficient $a(\tau)$ over the temporal support of the Green's function (over the ultrashort pulse duration) is negligible and can therefore be locally evaluated at a representative time $\tau$. Consequently, eq.~\eqref{Greens_function} provides an accurate description of the temporal diffraction of $\chi(\tau,z)$ over propagation distances for which the adiabatic condition remains valid, enabling a straightforward reconstruction of the full field dynamics.

With the knowledge of the Green's function $G_\chi$, it may result of interest to analyze Gaussian wave propagation and diffraction under temporal adiabatic modulations. For this, if we excite the optical system with an input $\tau_c$-centered Gaussian pulse defined by \linebreak $\chi_0(\tau) = \mathrm{exp}(-i(\tau-\tau_c)^2 / [2q_0])$, $q_0$  being a positive imaginary number that represents the initial Gaussian beam parameter, and then perform the convolution in eq. \eqref{convolution}, we obtain
\begin{equation}  \label{chi_solution}
    \chi(\tau, z) \approx  \sqrt{\frac{q_0}{q(\tau, z)}}\, \mathrm{exp}\left[ -i \frac{(\tau - \tau_{c})^2}{2q(\tau, z)} \right],
\end{equation}
where $q(\tau, z) = q_0 + 2a(\tau)z$. The time-dependent complex Gaussian parameter $q(\tau, z)$ encodes the properties of the temporal Gaussian pulse after propagation. As a difference with the definitions of $q$ in conventional spatial (time-invariant) paraxial optics, notice how now this parameter  evolves in time, although slowly under the adiabatic approximation. 

The time-dependence of $q(\tau, z)$ has subtle consequences. Early-time slices ($\tau$ small) may acquire slightly different temporal broadening and chirp compared to late-time slices ($\tau$ larger). In other words, each slice of the pulse evolves under a locally frozen dispersion coefficient $a(\tau)$, so that the propagation dynamics can be understood as a superposition of many quasi-independent Gaussian slices, each governed by its own effective complex temporal beam parameter. 

This temporal behavior is entirely analogous to a spatial Gaussian beam propagating through a lens with a slowly-varying refractive index across its transverse profile: in the spatial case, transverse slices of the beam encounter slightly different phase curvature, causing each slice to focus or diffract differently and producing a gradual overall modification of the beam profile without destroying its Gaussian nature. Similarly, in the temporal case, the pulse largely preserves its Gaussian envelope, but the local temporal width and chirp vary slightly across the pulse duration, reflecting the slow variation of the medium’s dispersion. As a result, the temporal evolution can be captured accurately by treating each slice independently using the local value of $a(\tau)$, with higher-order distortions remaining negligible within the adiabatic regime.

\begin{figure}[!t]
    \centering  \subfigure{\includegraphics[width=1\columnwidth]{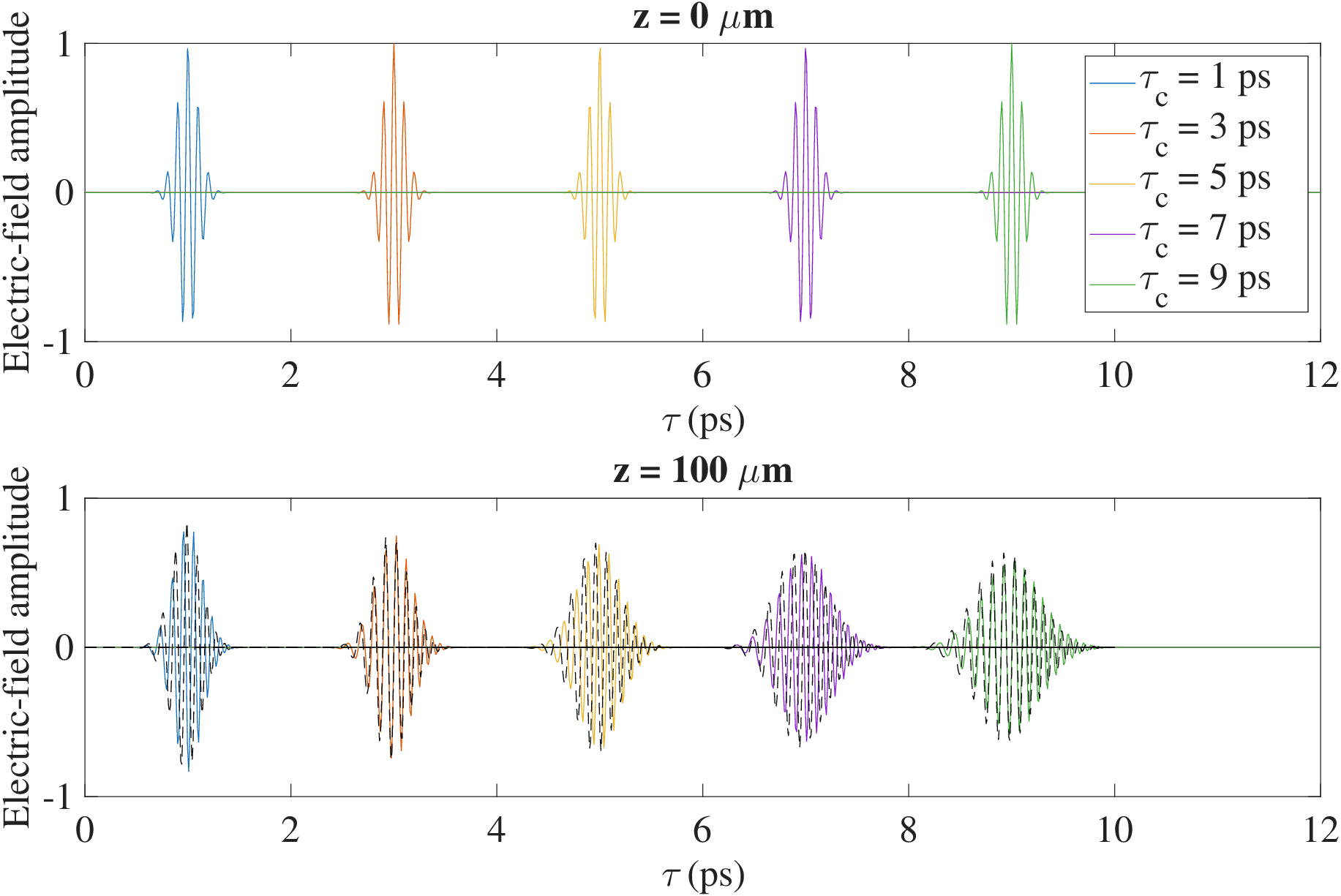}}
    \caption{Evolution of five identical pulses launched into a time-varying optical medium at five different time instants $\tau_\mathrm{c} =  1, 3, 5, 7, 9\,$ ps. (Top panel) Initial pulses at the spatial origin $z = 0$. (Bottom panel) Propagated and diffracted pulses at $z = 100\, \mu$m. System parameters: $q_{0} = i 10^{-26}\, \mathrm{s}^2$ , $\mu_{\text{r}}(t) = 1$, $\varepsilon_{\text{r}}(\tau) = \alpha \tau + 1.5$, $\alpha = 4\cdot 10^{11}\,\mathrm{s}^{-1}$. This permittivity definition gives $\varepsilon_{\text{r}}(1\,\mathrm{ps}) = 1.9$, $\varepsilon_{\text{r}}(3\,\mathrm{ps}) = 2.7$, $\varepsilon_{\text{r}}(5\,\mathrm{ps}) = 3.5$, $\varepsilon_{\text{r}}(7\,\mathrm{ps}) = 4.3$, $\varepsilon_{\text{r}}(9\,\mathrm{ps}) = 5.1\,$. Carrier frequency $f_\text{c} = 10\,$THz. Dashed black lines are results obtained numerically by FDTD.}
    \label{fig2}
\end{figure}

From a practical perspective, temporal adiabatic modulations open up interesting possibilities in optics, the most notable being the dynamical control of beam parameters with minimal distortion. To illustrate this, let us consider the scenario in \Fig{fig2}. There, five identical input Gaussian  pulses, 
$\chi_0 = \mathrm{exp}(-i(\tau-\tau_\mathrm{c})^2 / [2q_0])$, are launched into a time-modulated optical system at five different time instants $\tau_\mathrm{c}$, $\mathrm{c}=1,2,..,5$. Each pulse propagates essentially undistorted through the system, but because they locally experience different medium parameters $\varepsilon(\tau_\mathrm{c})$, their output beam widths and chirps differ, each characterized by $q_\mathrm{c}(\tau_\mathrm{c}, z)$. In the example, the dielectric constant of the medium is assumed to linearly vary in time according to $\varepsilon_{\text{r}}(\tau)~= \alpha \tau + 1.5$. The permeability is that of vacuum.

The top panel in \Fig{fig2} illustrates the pulses at the initial position $z = 0$, prior to propagation.  After propagating a distance of $100\, \mu$m (bottom panel), the differences in their temporal profiles become clearly visible.  Because each pulse samples the medium at a different launch time $\tau_c$, it experiences a distinct local dielectric response $\varepsilon(\tau_{\mathrm{c}})$, which in turn affects its temporal width and chirp.  The pulse launched later into the system (green curve) has widened more and its chirp has increased significantly compared to the pulse launched first (blue curve). The analytical results (solid lines) are supported by the numerical FDTD resolution of eq.~\eqref{wave_equation_time3} (dashed black lines).  This demonstrates that, by carefully choosing the launch time, one can dynamically tune the pulse’s temporal profile in a controlled manner while maintaining low distortion.  Notice that the primary difference between this numerical solution and the analytical one lies in the fact that $a(\tau)$ and $b(\tau)$ are no longer constant, but vary with $\tau$. The discrepancies arising from this time-dependence are practically negligible, thereby validating the analytical solution under the established conditions. Further tests were carried out by numerically solving Eq.~(5) under the assumption that $\partial^2 \psi/\partial z^2 \approx 0$. This numerical solution provides even greater accuracy and still predicts a remarkably similar pulse evolution. Although this second test falls outside the scope of the present work, it serves to reinforce the validity of the model under the applied approximations.

\begin{figure}[!t]
    \centering  
    \subfigure[]{\includegraphics[width=0.8\columnwidth]{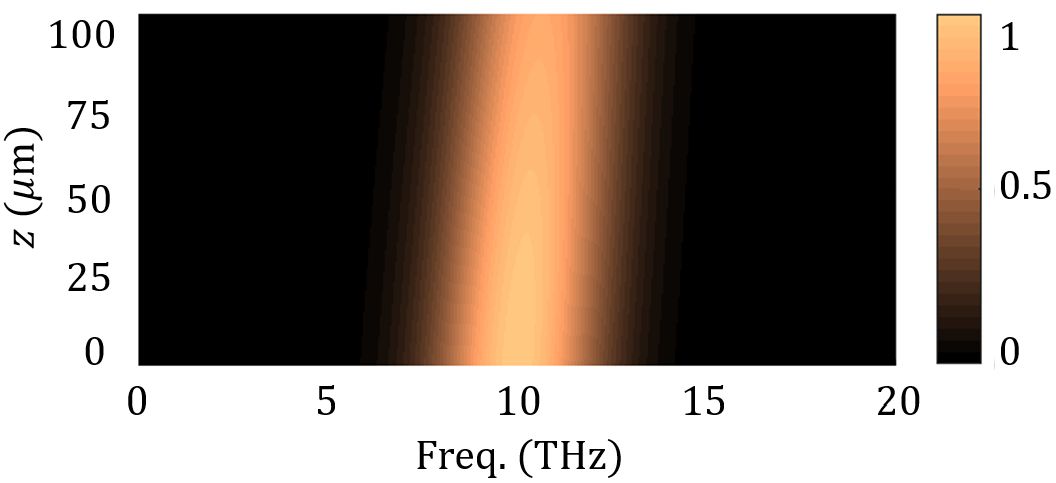}}
     \subfigure[]{\includegraphics[width=0.8\columnwidth]{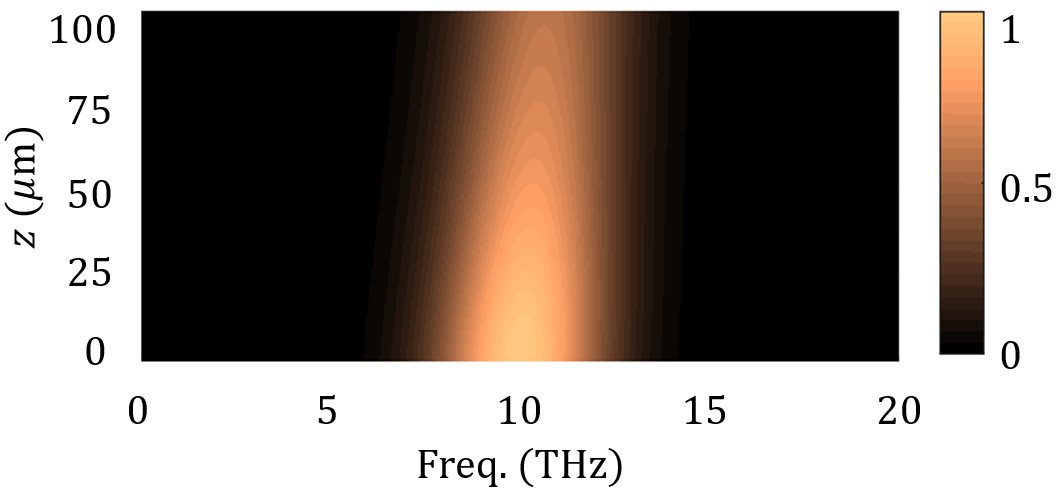}}
    \caption{Spectrum evolution along the propagation distance $z$ of the (a) first ($\tau_c = 1$ ps), and (b) last ($\tau_c = 9$ ps) pulses launched into the time-varying optical medium in Fig. 2.} 
    \label{fig3}
\end{figure}

Fig.~\ref{fig3} shows the spectral evolution  for the first (blue, $\tau_c = 1$ ps) and last (green, $\tau_c = 9$ ps) launched pulses as they propagate in $z$. The system parameters are the same as in Fig.~\ref{fig2}. As can be seen, the two pulses, which initially share the same carrier frequency (10 THz), evolve differently as a function of $z$. At $z = 100,\mu$m, the pulse centered at 1 ps preserves a broader spectral content than the pulse centered at 9 ps. This is consistent with the results shown in Fig.~\ref{fig2}, since in the time domain the 1-ps-centered pulse is narrower than the 9-ps-centered pulse. In addition, it is worth noting that the center (carrier) frequency of the pulses has slightly changed after propagation. This frequency shift, appreciable at $z = 100\, \mu$m, is a direct consequence of the time-dependent parameters in the time-modulated media, which break time-translation symmetry and allow frequency conversion. For the selected modulation in the example, the center (carrier) frequency progressively increases as the propagation distance increases too. However, the magnitude and direction of the frequency shift depend on the specific temporal modulation profile and may differ for other modulations.

Unlike time-varying approaches that rely on time-modulated conductive thin sheets like indium tin oxide (ITO), the theoretical framework explored in this work is tailored for bulk media. Towards a feasible and practical implementation, macroscopic bulk materials such as liquid crystals can be used to mimic an adiabatically time-modulated homogeneous medium, where their millisecond response times appear slow and quasi-stationary relative to ultrafast electromagnetic pulses \cite{khoo2014liquid}. Alternatively, if faster temporal switching is required, solid-state electro-optic materials such as Lithium Niobate ($\text{LiNbO}_3$) can be explored \cite{Boesetal2023}. Using these materials, one can envision a waveguide device filled with a reconfigurable, time-modulated bulk medium. When illuminated by a sequence of Gaussian pulses launched at different times, each pulse encounters a different state of the time-modulated environment. Consequently, the pulses experience varying amounts of diffraction and chirping, enabling dynamic wave manipulation.

\section{Adiabatic ABCD Matrix Formalism}

In the temporal paraxial scenario, we aim to find a matrix relation that connects $(\tau, \Omega)$ with $(\tau_0, \Omega_0)$ in a way that
\begin{equation} \label{ABCD}
    \begin{pmatrix}
    \tau  \\
    \Omega  
\end{pmatrix}
    =
    \begin{pmatrix}
    A & B \\
    C & D 
\end{pmatrix}
\begin{pmatrix}
    \tau_0\\
    \Omega_0  
\end{pmatrix},
\end{equation}
analogous to the well-known ABCD matrix formalism in spatial paraxial optics, where the transverse position and angle at the output plane $(y, \theta)$ are related to these parameters at the input plane $(y_0, \theta_0)$  \cite{GoodmanBook, wolf2004geometric, Kogelnik66, Belanger91}.  

The utility of this approach lies in its ability to describe the evolution of the pulse envelope in terms of linear transformations, capturing effects such as temporal spreading, chirping, and compression in a manner fully analogous to beam diffraction, focusing, and lensing in spatial optics. This has already been exploited in other contexts, such as plane-wave propagation in spatial and temporal dielectric composites and photonic time crystals \cite{Molero2025}. The ABCD matrix formalism can also be combined with the temporal Green's function or Fourier-transform methods to compute the full wave-packet evolution in both time and frequency domains, making it a powerful tool for analyzing ultrashort pulse dynamics in temporally varying optical systems. 

In the temporal context where the ABCD relation \eqref{ABCD} is defined, $\tau_0$ and $\Omega_0$ denote the initial temporal coordinate and frequency detuning of a pulse at the source plane, while $\tau$ and $\Omega$  describe their corresponding values after propagation through the time-modulated medium. The parameter $A$ relates transverse temporal magnification, $B$ governs pulse broadening and temporal diffraction, $C$ describes time-lensing phenomena, and $D$ controls frequency scaling.

In the following, we derive the ABCD matrix of temporal Gaussian pulse propagation in adiabatically-modulated media from Hamiltonian principles, thus connecting the topic to the concepts depicted in Section III. Nonetheless, it should be stated that this would represent a particular propagation scenario. The ABCD framework is of more general nature and, in principle, interesting practical phenomena such as temporal lensing \cite{Kolner1994} admit to be analyzed and extended from time-invariant to time-varying contexts.   

After defining the frequency operator $\hat{\Omega} = i \partial / \partial \tau$, the time-dependent Hamiltonian of the reduced temporal paraxial equation \eqref{wave_equation_time4} is expressed as $\hat{H}_\chi = a(\tau) \hat{\Omega}^2 $. Thus, Hamilton's equations give
\begin{equation}
    \dot{\tau} = \frac{\partial H_\chi}{\partial \Omega} = 2a(\tau) \Omega,
\end{equation}
\begin{equation}
    \dot{\Omega} = - \frac{\partial H_\chi}{\partial \tau} = -\Omega^2  \frac{\partial a(\tau)}{\partial \tau}   \approx 0.
\end{equation}
When $a$ is constant (the medium is not modulated in time), $\dot{\Omega}$ is strictly zero. This implies that there is no frequency conversion, latter observed in the relation $\Omega = \Omega_0$. In contrast, when $a$ varies in time, $\dot{\Omega}$ is generally nonzero and depends quadratically on $\Omega$, so that the frequency evolution is intrinsically nonlinear and a simple linear relation between $\Omega$ and $\tau$ cannot be defined.
In the adiabatic regime, where the pulse duration is much shorter than the characteristic modulation period, the temporal variation of $a(\tau)$ can be neglected locally, leading to $\dot{\Omega}\approx 0$. In this case, the central frequency remains approximately conserved over short propagation distances, consistent with the \emph{locally} quasi-static character of the medium.

Integration of $\dot{\tau}$ and $\dot{\Omega}$ along the evolution variable $z$ leads to $\tau = \tau_0+ 2a(\tau) z \Omega_0$ and $\Omega = \Omega_0$, which can be rearranged into matrix form as
\begin{equation}
    \begin{pmatrix}
    \tau  \\
    \Omega  
\end{pmatrix}
    =
    \begin{pmatrix}
    1 & 2a(\tau)z \\
    0 & 1 
\end{pmatrix}
\begin{pmatrix}
    \tau_0\\
    \Omega_0  
\end{pmatrix},
\end{equation}
The ABCD parameters that describe Gaussian beam propagation in the adiabatically-modulated temporal medium are therefore $A = 1$, $B= 2a(\tau)z$,  $ C = 0, D = 1$. The fact that $B = B(\tau)$ connects directly with the time-dependent complex Gaussian parameter $q = q(\tau, z)$ derived above. It denotes that pulses entering the medium at different times experience different temporal diffraction, allowing systematic control of pulse characteristics across sequential pulses. 

In addition, notice that the determinant $AD - BC = 1$. The unimodularity of the ABCD matrix reveals the conservation of phase-space area in paraxial optics. In the spatial case, it implies that the optical system is reversible and lossless. As a consequence, the \emph{étendue}, defined as the area in the $(y,\theta)$ phase space, is conserved during propagation. This reflects Liouville’s theorem, according to which the phase-space measure is preserved under Hamiltonian evolution \cite{GerrardBurch1994}. These same properties can be translated to the temporal case, where the  $(\tau, \Omega)$ defines an equivalent frequency-temporal phase space, establishing its conservation as a fundamental invariant.

The quadratic phase appearing in the adiabatic Green's function~\eqref{Greens_function} can be directly recovered from the classical action \(S\). Defining the action in the Hamiltonian sense as
\begin{equation}
S = \int_0^z \left( \Omega \dot{\tau} - H \right)\, \mathrm{d}z ,
\end{equation}
and substituting the expressions for \( \Omega \), \( \dot{\tau} \), and \( H \), one obtains
\begin{equation}
S = \frac{(\tau - \tau_0)^2}{4 a(\tau) z} ,
\end{equation}
which coincides with the exponent of $G_\chi$ in eq.~\eqref{Greens_function} (Supplementary Material for details). This reveals the underlying Hamilton--Jacobi structure of  paraxial dynamics in adiabatically time-modulated media, identifying the phase of the Green's function with the classical action of temporal ray trajectories. The correspondence is directly analogous to spatial paraxial optics and free-particle quantum mechanics.

The matrix formalism is also highly effective for estimating and quantifying both the chirp and width of the Gaussian pulse. This direct connection between $q(\tau, z)$ and the matrix coefficients is particularly useful for addressing inverse problems, where achieving a target chirp and width is the ultimate goal. Consequently, this allows for the determination of the appropriate matrix coefficients, which, in the present case, simplifies to the strategic manipulation of the parameter $B = +2a(\tau)z$.

Accordingly, the width $w$ and chirp $K$ of a $\tau_c$-centered temporal Gaussian pulse that has propagated a distance $z$ can be approximated via the following identity, $\text{exp} [-i \tau^2 / (2q)] = \text{exp} [- \tau^2 (1+iK) / (2w^2)] $, which leads to 
\begin{equation}
\label{width} w = \sqrt{\frac{|q_{0}|^2 + 4a^2(\tau_{\text{c}}) z^2}{|q_{0}|}}, 
\end{equation}
\begin{equation}
\label{chirp}
K =  \frac{2 a(\tau_{c}) z}{|q_0|} \,.
\end{equation}
Here, the parameter $w$ physically represents the pulse half-width at which the electric field amplitude drops to $e^{-1/2} \approx 0.607$ of its peak value. Within the framework of the adiabatic approximation, these two parameters are evaluated at the local temporal center of the pulse.

\section{Conclusion}
In this work, we have analyzed the  dynamics of ultrashort temporal pulses in slowly-varying (adiabatic) time-modulated media. In analogy to conventional spatial paraxial optics, we have derived a Schrödinger-like  wave equation that accounts for wave propagation and diffraction of temporal pulses. Green's-function and Hamiltonian methods are invoked to obtain an approximate, but analytical, adiabatic solution, which also admits an ABCD matrix representation. Results show how the pulse features (width, chirp) can be dynamically  tailored without significant distortion at short distances by simply leveraging temporal modulations, not just with the conventional medium dispersion present in time-invariant materials. This is expected to open up new design possibilities in the control of ultrashort optical pulses and have direct implications in the fields of time-modulated photonics and metamaterials. 
 
\section*{Acknowledgments}

We thank Prof. José Capmany for the initial discussions on the subject and for pointing us to relevant state-of-the-art literature. 

This work was supported in part by the BBVA Foundation's Leonardo Grant for Scientific Research and Cultural Creation 2025, in part by the Grant PID2024-155167OA-I00 funded by MICIU/AEI/10.13039/501100011033/FEDER, UE, and in part by Consejería de Universidad, Investigación e Innovación of Junta de Andalucía through grant EMERGIA 23-00235. The BBVA Foundation is not responsible for the opinions, comments, and content included in the project and/or the results derived from it, which are the sole and absolute responsibility of their authors.

\section*{Disclosures}
The authors declare no conflicts of interest.

\section*{Data Availability}
Data underlying the results presented in this paper are not publicly available at this time but may be obtained from the authors upon reasonable request.


\bibliography{ref}



\pagebreak

\setcounter{equation}{0}
\setcounter{figure}{0}
\setcounter{table}{0}
\setcounter{page}{1}
\makeatletter
\renewcommand{\theequation}{S\arabic{equation}}
\renewcommand{\thefigure}{Supp.\arabic{figure}}
\renewcommand{\labelenumi}{\arabic{enumi}.}
\renewcommand{\thetable}{Supp.\arabic{table}}

\newgeometry{left=1.5cm,right=1.5cm}

\onecolumn
\begin{center}
\textrm{\huge Supplementary Material}
\end{center}

\vspace{0.2cm}

\begin{center}
\textrm{\LARGE Temporal Paraxial Optics under Adiabatic Modulations}
\end{center}

\begin{center}
\textrm{\Large Antonio Alex-Amor$^{*}$, Carlos Molero$^\ddagger$}
\end{center}

\begin{center}
\textit{$*$Department of Electronic and Communication Technology, RFCAS Research Group,  Universidad Autónoma de Madrid, \\ 28049 Madrid, Spain.}
\end{center}

\begin{center}
\textit{$\ddagger$Department of Electronic and Electromagnetism, Faculty of Physics, University of Seville, 41012, Seville, Spain.}
\end{center}

\section*{Development of the General Temporal Wave Equation with an ansatz of the form $\psi(t,z)\, \mathrm{e}^{-i\omega_0 t} \mathrm{e}^{+ik_0 z}$}

In this section, we show how  the general temporal scalar wave equation \eqref{scalar_waveequation}   expands into \eqref{wave_equation_time1} when an ansatz  $E_x = \psi(t,z)\, \mathrm{e}^{-i\omega_0 t} \mathrm{e}^{+ik_0 z}$ is considered. 

First, we compute the second spatial derivative (left-hand side):
\begin{equation} \label{lefthand}
    \begin{aligned}
    \frac{\partial ^2 E_x}{\partial z^2} 
    &=
    \frac{\partial}{ \partial z} \left[ \frac{\partial (\psi\, \mathrm{e}^{-i\omega_0 t} \mathrm{e}^{+ik_0 z})}{\partial z} \right] 
    =
    \mathrm{e}^{-i\omega_0 t} \frac{\partial}{ \partial z} \left[ \frac{\partial (\psi\,  \mathrm{e}^{+ik_0 z})}{\partial z} \right] 
    = \mathrm{e}^{-i\omega_0 t} \frac{\partial}{ \partial z} \left[ \frac{\partial \psi}{\partial z} 
     \mathrm{e}^{+ik_0 z} + ik_0 \psi \mathrm{e}^{+ik_0 z} \right] \\
     & = 
     \mathrm{e}^{-i\omega_0 t} \mathrm{e}^{+ik_0 z} \left[ \frac{\partial ^2 \psi}{\partial z^2} + 2ik_0 \frac{\partial \psi}{\partial z} - k_0^2 \psi   \right]
    \end{aligned}
\end{equation}

Now, we compute the right-hand side (temporal derivatives). We first start by computing,
\begin{equation}
    \begin{aligned}
    \frac{\partial (\varepsilon E_x)}{\partial t} 
    &= 
    \mathrm{e}^{+ik_0 z}\,
    \frac{\partial }{\partial t} \left[ \varepsilon \psi\, \mathrm{e}^{-i\omega_0 t} \right]
    = 
    \mathrm{e}^{+ik_0 z} \left[
    \frac{\partial \varepsilon}{\partial t} \psi \mathrm{e}^{-i\omega_0 t}  +  \varepsilon \frac{\partial (\psi \mathrm{e}^{-i\omega_0 t}) }{\partial t}  
    \right]
    = 
    \mathrm{e}^{-i\omega_0 t} \mathrm{e}^{+ik_0 z} \left[
     \frac{\partial \varepsilon}{\partial t} \psi + \varepsilon \frac{\partial \psi}{\partial t} - i\omega_0 \varepsilon \psi 
    \right]\, ,
    \end{aligned}
\end{equation}
and then compute the full right-hand side (the full time derivative term):
\begin{equation} \label{righthand1}
    \begin{aligned}
    \frac{\partial}{\partial t} \left[\mu \frac{\partial (\varepsilon E_x)}{\partial t} \right] 
    &= 
    \mathrm{e}^{+ik_0 z} \frac{\partial}{\partial t} \left[  \mu \frac{\partial \varepsilon}{\partial t} \psi \mathrm{e}^{-i\omega_0 t}  + \mu \varepsilon \frac{\partial \psi} {\partial t}\mathrm{e}^{-i\omega_0 t}  - i\omega_0 \mu \varepsilon \psi \mathrm{e}^{-i\omega_0 t}   \right]
    =\mathrm{e}^{+ik_0 z} \big[ \partial t_1 + \partial t_2 + \partial t_3 
    \big],  
    \end{aligned}
\end{equation}
where $\partial t_i$ ($i=1,2,3$) are auxiliary terms, representing the three addends, that simplify the calculation. These terms are
\begin{equation}
    \begin{aligned}
    \partial t_1 = \frac{\partial}{\partial t} \left[ \mu \frac{\partial \varepsilon}{\partial t} \psi \mathrm{e}^{-i\omega_0 t} \right] 
    =
    \mathrm{e}^{-i\omega_0 t} \left[ \frac{\partial \mu}{\partial t}  \frac{\partial \varepsilon}{\partial t} \psi + \mu \frac{\partial^2 \varepsilon}{\partial t^2} \psi + \mu \frac{\partial \varepsilon}{\partial t} \frac{\partial \psi}{\partial t} - i\omega_0  \mu \frac{\partial \varepsilon}{\partial t} \psi 
    \right],
    \end{aligned}
\end{equation}
\begin{equation}
    \begin{aligned}
    \partial t_2 = \frac{\partial}{\partial t} \left[\mu \varepsilon \frac{\partial \psi} {\partial t}\mathrm{e}^{-i\omega_0 t} \right] 
    =
    \mathrm{e}^{-i\omega_0 t} \left[ \varepsilon \frac{\partial \mu}{\partial t}  \frac{\partial \psi}{\partial t} 
    +
    \mu \frac{\partial \varepsilon}{\partial t}  \frac{\partial \psi}{\partial t}
    +
    \mu \varepsilon \frac{\partial^2 \psi}{\partial t^2}
    -
    i\omega_0 \mu \varepsilon \frac{\partial \psi}{\partial t}
    \right],
    \end{aligned}
\end{equation}
\begin{equation}
    \begin{aligned}
    \partial t_3 = \frac{\partial}{\partial t} \left[-i\omega_0 \mu \varepsilon \psi \mathrm{e}^{-i\omega_0 t} \right] 
    = -i \omega_0 \mathrm{e}^{-i\omega_0 t} \left[ \mu \frac{\partial \varepsilon}{\partial t}  \psi +
    \varepsilon \frac{\partial \mu}{\partial t} \psi + 
    \varepsilon \mu \frac{\partial \psi}{\partial t} - 
    i\omega_0 \varepsilon \mu \psi
    \right]\, .    
    \end{aligned}
\end{equation}
After inserting the values of $\partial t_i$ ($i = 1,2,3$) into eq. \eqref{righthand1} and reordering terms, we get
\begin{equation} \label{righthand2}
    \frac{\partial}{\partial t} \left[\mu \frac{\partial (\varepsilon E_x)}{\partial t} \right] 
    =
    \mathrm{e}^{-i\omega_0 t} \mathrm{e}^{+ik_0 z} \, \left[ \mu \varepsilon \left(\frac{\partial^2 \psi}{\partial t^2} - i2\omega_0 \frac{\partial \psi}{\partial t} - \omega_0^2 \psi \right)  
    + 
    \left(\frac{\partial \psi}{\partial t} - i\omega_0 \psi \right) \left( \varepsilon \frac{\partial \mu}{\partial t} + 2\mu \frac{\partial \varepsilon}{\partial t} \right) 
     + 
    \left(\mu \frac{\partial^2 \varepsilon}{\partial t^2}+ \frac{\partial \mu}{\partial t} \frac{\partial \varepsilon}{\partial t} \right) \psi \right] 
\end{equation}

Eqs.~\eqref{lefthand} and \eqref{righthand2} are the left-hand and right-hand sides of eq.~\eqref{wave_equation_time1}, respectively. The exponential functions $\mathrm{e}^{-i\omega_0 t} \mathrm{e}^{+ik_0 z}$ cancel out, leading to the general temporal scalar wave equation \eqref{wave_equation_time1} presented in the main manuscript.

\section*{Paraxial Optics with Spatial Modulations}

Here, we complete the derivation of the spatial paraxial wave equation \eqref{wave_equation_space} in the main manuscript when spatial modulations along the transverse direction $y$ are considered. 

We consider linear, isotropic and non-dispersive media of the form 
\begin{equation}
    \mathbf{D}(y,z) = \varepsilon(y) \mathbf{E}(y,z), \quad \quad
    \mathbf{B}(y,z) = \mu(y) \mathbf{H}(y,z) 
\end{equation}
The temporal dependence is time-harmonic, $\mathrm{e}^{- i\omega_0 t}$, and is therefore omitted. Moreover, we assume linearly-polarized fields of the form
\begin{equation}
    \mathbf{E} = \hat{\mathbf{x}}E_x, \quad \quad \mathbf{H} = \hat{\mathbf{y}}H_y,
\end{equation}

In this scenario, the spectral scalar wave equation for $E_x$, derived directly from Maxwell's equations, reads
\begin{equation} \label{wave_equation_space_sup}
    \frac{\partial^2 E_x}{\partial y^2}  + \frac{\partial^2 E_x}{\partial z^2} = - \frac{\omega_0^2}{v^2} E_x,
\end{equation}
with $v^2 = 1/(\varepsilon \mu)$ the propagation velocity in the medium. For the sake of readability, we drop the $y$ dependence of $\varepsilon(y)$, $\mu(y)$ and $v(y)$. 

Now, we assume an ansatz of the form
\begin{equation} \label{Ex_space}
    E_x(y,z) = \psi(y,z) \mathrm{e}^{+ik_0 z},
\end{equation}
The former expression essentially represents a  field envelope $\psi$ modulated by a plane wave term $\mathrm{e}^{+ik_0 z}$. It can serve to model a spatial Gaussian beam.

The insertion of the Ansatz \eqref{Ex_space} into the wave equation \eqref{wave_equation_space_sup} leads to
\begin{equation} \label{wave_equation_space2}
    \frac{\partial^2 \psi}{\partial y^2}  + \frac{\partial^2 \psi}{\partial z^2} + i2k_0 \frac{\partial \psi}{\partial z} = \left[k_0^2  - \frac{\omega_0^2}{v^2} \right] \psi
\end{equation}

In the \emph{paraxial regime} ($\psi$ is a slowly-varying function along the propagation direction $z$), $|\partial_z^2 \psi| \ll |k_0\cdot  \partial_z \psi|$. Thus, the term $\partial_z^2 \psi$ can be neglected. 

We can simplify \eqref{wave_equation_space2} as
\begin{equation} \label{wave_equation_space_paraxial}
   \frac{\partial^2 \psi}{\partial y^2}  
    + i2k_0 \frac{\partial \psi}{\partial z} \approx \left[k_0^2  - \frac{\omega_0^2}{v^2} \right] \psi
\end{equation}
Dividing \eqref{wave_equation_space_paraxial} by $+i2k_0$ gives us the spatial paraxial equation
\begin{equation} \label{wave_equation_space_paraxial2}
    \frac{\partial \psi(y,z)}{\partial z} \approx +\frac{i}{2k_0} \frac{\partial^2 \psi(y,z)}{\partial y^2} - \frac{i \omega_0}{2c} \left(1 - \frac{c^2}{v^2(y)} \right) \psi(y,z),
\end{equation}
with $c = \omega_0 / k_0$ is the speed of light in vacuum.

\section*{Derivation of the Temporal Adiabatic Green's Function}

In order to derive the temporal adiabatic Green's function \eqref{Greens_function} in the main manuscript, we follow these steps. The main underlying idea is to consider that the values of the time-dependent coefficients $a(\tau)$ and $b(\tau)$  do not change much during the duration of the ultra-short pulse; namely, $a(\tau)$ and $b(\tau)$ are practically constant in the interval $[\tau - T_\mathrm{pulse}/2, \tau + T_\mathrm{pulse}/2]$, $T_\mathrm{pulse}$ being the duration of the pulse. Thus, to all practical effect, we consider $a(\tau) = a_0$ and $b(\tau) = b_0$ constant terms.

Then, we define the Fourier relations $\tilde{\chi}(\Omega, z) = \int_{-\infty}^\infty \chi(\tau,z) \mathrm{e}^{+i\Omega  \tau}\, d\tau$ and $\chi(\tau, z) = \frac{1}{2\pi} \int_{-\infty}^\infty \tilde{\chi}(\Omega,z) \mathrm{e}^{-i\Omega  \tau}\, d\Omega$. Since $\partial_\tau^2 \chi \rightarrow -\Omega^2 \tilde{\chi}$, we express the reduced temporal wave equation \eqref{wave_equation_time4} in the Fourier domain as
\begin{equation}
    \frac{\partial \tilde{\chi}}{\partial z} \approx + ia_0\Omega^2 \tilde{\chi},
\end{equation}
whose solution is
\begin{equation}
    \tilde{\chi}(\Omega, z) = \tilde{\chi}_0(\Omega) \mathrm{e}^{+ia_0 \Omega^2 z}.
\end{equation}
The term $\tilde{\chi}_0(\Omega)$ is the initial condition in the frequency domain. The time-domain initial condition that leads to the Green's function is $\chi_0(\tau, \tau_0) = \delta (\tau - \tau_0)$, which corresponds to $\tilde{\chi}_0(\Omega) = \mathrm{e}^{+i\Omega  \tau_0}$ after Fourier transformation. 

Thus, the Green's function $G_\chi(\tau, z; \tau_0)$ is computed as
\begin{equation}
    G_\chi(\tau, z; \tau_0) = \frac{1}{2\pi}  \int_{-\infty}^\infty \tilde{\chi}(\Omega,z) \mathrm{e}^{-i\Omega  \tau}\, d\Omega = 
    \frac{1}{2\pi} \int_{-\infty}^\infty \mathrm{e}^{+i\Omega  \tau_0} \mathrm{e}^{+ia_0 \Omega^2 z} \mathrm{e}^{-i\Omega  \tau}\, d\Omega = 
    \frac{1}{2\pi} \int_{-\infty}^\infty \mathrm{e}^{-i\Omega (\tau- \tau_0)} \mathrm{e}^{+ia_0 \Omega^2 z} \, d\Omega.
\end{equation}
The above is a gaussian integral. To solve it, we use\footnote{Strictly, this identity is well defined for $\mathrm{Re}\{A \}>0$. As seen, $A$ is imaginary in our problem. Therefore, we intrinsically assume $A = \delta - ia_0 z$, $\delta$ being a positive real number, and then compute  $\delta \rightarrow 0^+$. This assumes that the oscillations cancel at infinity.}
\begin{equation} \label{identity}
    \int_{-\infty}^\infty \mathrm{exp}\big[Bx -Ax^2  \big]\, dx = \sqrt{\frac{\pi}{A}}\, \mathrm{exp}\left[ \frac{B^2}{4A} \right], 
\end{equation}
where we identify $B = -i(\tau-\tau_0)$ and $A = -ia_0z$. Then, the Green's function reads
\begin{equation}
     G_\chi(\tau, z; \tau_0) = \frac{1}{\sqrt{-i4 \pi a_0 z}} 
\exp\Bigg[ -i \frac{(\tau-\tau_0)^2}{4 a_0 z} \Bigg]\, .
\end{equation}
This solution is exact if $a_0$ is constant.  In the adiabatic regime, we may approach $a_0 \approx a(\tau)$, which gives  eq. \eqref{Greens_function} in the main manuscript. 

\section*{Derivation of the Temporal Gaussian Adiabatic Solution for $\chi(\tau, z)$}

When an input Gaussian pulse $\chi_0(\tau) = \mathrm{exp}[-i\tau^2 / (2q_0)]$ excites the optical system, the output $\chi(\tau, z)$ is computed via the convolution in eq. \eqref{convolution}. That is,
\begin{equation}
    \chi(\tau, z) = \int_{-\infty}^\infty G_\chi(\tau, z; \tau_0)\, \chi_0(\tau_0)\, d\tau_0 = 
    \int_{-\infty}^\infty \frac{1}{\sqrt{-i4 \pi a_0 z}} 
\exp\Bigg[- i \frac{(\tau-\tau_0)^2}{4 a_0 z}\Bigg]\, \mathrm{exp}\left[-i\frac{\tau_0^2}{2q_0}\right]\, d\tau_0 \, .
\end{equation}

By expanding the quadratic term in the Green's function exponent, the equation can be rearranged as\begin{align}\chi(\tau, z) &=  \frac{1}{\sqrt{-i4 \pi a_0 z}}\mathrm{exp} \Bigg[ -i\frac{\tau^2}{4a_0z} \Bigg] \int_{-\infty}^\infty\mathrm{exp} \Bigg[ +i\frac{2\tau}{4a_0z}\tau_0 - i\left( \frac{1}{2q_0} + \frac{1}{4a_0 z} \right) \tau_0^2 \Bigg]\, d\tau_0.\end{align}
We can evaluate this using the standard Gaussian integral identity in eq. \eqref{identity}. By identifying the parameters as $B = +i2\tau / (4a_0z)$ and $A = i [1/(2q_0) + 1/(4a_0 z)]$, the right-hand side integration yields\begin{equation}\chi(\tau, z) = \sqrt{\frac{q_0}{q_0 + 2a_0 z}} \mathrm{exp} \Bigg[- i\frac{\tau^2}{4a_0z} \Bigg]  \mathrm{exp} \Bigg[ +i\frac{\tau^2 q_0}{4a_0z (q_0 + 2a_0z)} \Bigg].\end{equation}
After merging the two exponents and defining the standard complex temporal Gaussian beam parameter $q = q_0 + 2a_0z$, one obtains the final expression for $\chi$:\begin{equation} \label{chi_origin_supp}\chi(\tau, z) = \sqrt{\frac{q_0}{q}}\,  \mathrm{exp} \Bigg[ -i\frac{\tau^2}{2q} \Bigg].\end{equation}This solution is exact for a constant term $a(\tau) = a_0$. In the constant case, $q = q(z)$ would be only a function of the evolution spatial variable $z$, perfectly mirroring the standard $q$-parameter propagation found in paraxial spatial optics. In the time-dependent case, we can extend this solution in the adiabatic regime by assuming $a(\tau) \approx a_0$ during the duration of the pulse. This matches the solution reported in eq. \eqref{chi_solution} of the main manuscript for an initial pulse centered at the origin. If the pulse is centered at a generic instant $\tau_c$, we simply replace $\tau \rightarrow \tau-\tau_c$ in eq. \eqref{chi_origin_supp}.

\section*{Connection between Action $S$, Hamiltonian $H$ and Green's Function $G_\chi$}

Taking into account that the evolution variable is $z$, the action $S$ is defined in the classical sense as 
\begin{equation} \label{actionS}
    S = \int_0^z (\Omega \dot{\tau} - H) \, dz\, .
\end{equation}
The values of the Hamiltonian $H = a(\tau) \Omega^2$, and the spectral terms $\dot{\tau} = 2a(\tau) \Omega$ and $\Omega \approx \Omega_0$, are derived in Section 4 of the main manuscript for the temporal adiabatic scenario. Moreover, in the adiabatic regime, the temporal variation of the medium is locally quasi-static during propagation, allowing us to approximate $a(\tau) \approx a_0$ along the ray trajectory over the distance $z$. Substitution of these values into eq. \eqref{actionS} gives
\begin{equation} \label{actionS2}
    S =\int_0^z (2a(\tau) \Omega_0^2 - a(\tau) \Omega_0^2)\, dz = \int_0^z a(\tau) \Omega_0^2 \, dz = a(\tau) z \Omega_0^2
\end{equation}
From the main manuscript, we also know that $\tau = \tau_0 + 2a(\tau) z \Omega_0$. Solving for $\Omega_0$, we find
\begin{equation}
    \Omega_0 = \frac{\tau - \tau_0}{2a(\tau) z}\, .
\end{equation}
Substitution of the value of $\Omega_0$ into eq. \eqref{actionS2} leads to an action $S$ of the form
\begin{equation} \label{actionS3}
    S = a(\tau) z \Omega_0^2 = \frac{(\tau - \tau_0)^2}{4a(\tau) z}\, .
\end{equation}

As shown above and in the main manuscript [eq. \eqref{Greens_function}], the Green's function $G_\chi \propto 
\exp\left[ -i \frac{(\tau-\tau_0)^2}{4 a(\tau) z} \right] $. We readily notice that the action $S$ in eq. \eqref{actionS3} is linked to the exponent (phase) of the Green's function. This correspondence reflects the underlying Hamilton–Jacobi structure of paraxial temporal dynamics in adiabatic time-modulated media. Therefore, the Green’s function phase emerges naturally as the classical action associated with temporal ray trajectories, in direct analogy with spatial paraxial optics and free-particle quantum mechanics. This result provides a clear physical interpretation of temporal diffraction and dispersion as a classical propagation process governed by the action accumulated along adiabatic temporal rays.

\section*{Evaluation of the robustness of the Analytical Gaussian Waveform Solution}
\noindent The temporal paraxial approximation relies on two key assumptions: a slowly-varying envelope along $z$, such that $\partial^2 \Psi / \partial z^2 \approx 0$, and a temporally slowly-varying permittivity and permeability. The latter motivates the adiabatic regime, where permittivity and permeability remain locally constant during fast pulse propagation, where $v(t)$ and later $a(\tau), b(\tau)$ are assumed fixed. We evaluate and compare the solutions to these equations under their respective approximations:
\begin{itemize}
\item Eq.~(3) in the main manuscript is the full wave equation under the ansatz $E_{\text{x}}(t, z) = \psi(t, z) \text{e}^{-i \omega_{0} t} \text{e}^{+ i k_{0}z}$. 
Enforcing $\partial^2 \psi(t, z)/\partial z^2 \approx 0$ yields a numerical solution. The inclusion of the variable change $t = \tau + zc/v(\tau)$ is realized, in such a way that $\psi(t, z) \rightarrow \psi(\tau, z)$, thus forcing
\begin{align}
\frac{\partial \psi(t, z)}{\partial t} &= \frac{\partial \psi(\tau, z)}{\partial \tau} \frac{\partial \tau}{ \partial t} \hspace{1mm}\text{with} \hspace{1mm} \frac{\partial \tau}{\partial t} = \bigg[\frac{\partial t}{\partial \tau}\bigg]^{-1} = \frac{1}{1 + (z / c)[\partial \epsilon_{\text{r}}(\tau)/\partial \tau]} \\
\frac{\partial \psi(t, z)}{\partial z} &= \frac{\partial \psi(\tau, z)}{\partial z} - \frac{1}{1 + (z/c)[\partial \varepsilon_{\text{r}(\tau)}/ \partial \tau]}\frac{\partial \psi(\tau, z)}{\partial \tau} \frac{\partial \tau}{\partial z} \hspace{1mm} \text{with} \hspace{1mm} \frac{\partial \tau}{\partial z} =  \frac{\varepsilon_{\text{r}}(\tau)}{c} 
\end{align}
where we have used $v^2(\tau) = \frac{1}{\varepsilon_{\text{r}}(\tau) \mu_{\text{r}}}$ ($\mu_{\text{r}} = 1$). Time derivatives concerning $\varepsilon_{\text{r}}(\tau)$ are not neglected.
\item Eq.~(5) in the main manuscript assumes $\partial\varepsilon / \partial t = \partial\mu/\partial t \approx 0$. Introducing the transformed time variable $\tau = t - zc/v^2$ leads to a numerical solution as well. Times derivatives of $\varepsilon_{\text{r}}(\tau)$ are neglected.
\item Assuming that $a(\tau)$ and $b(\tau)$ vary slowly and can be treated as approximately constant terms,  Eq.~(5) reduces to the analytical solution given in Eq.~(13) of the main manuscript.
\end{itemize}
In terms of approximation level, Eq.~(13) is the most approximated, followed by the numerical solution of Eq.~(5), and the numerical solution of Eq.~(3).  \\

\begin{figure}[!h]
    \centering      \includegraphics[width=0.8\columnwidth]{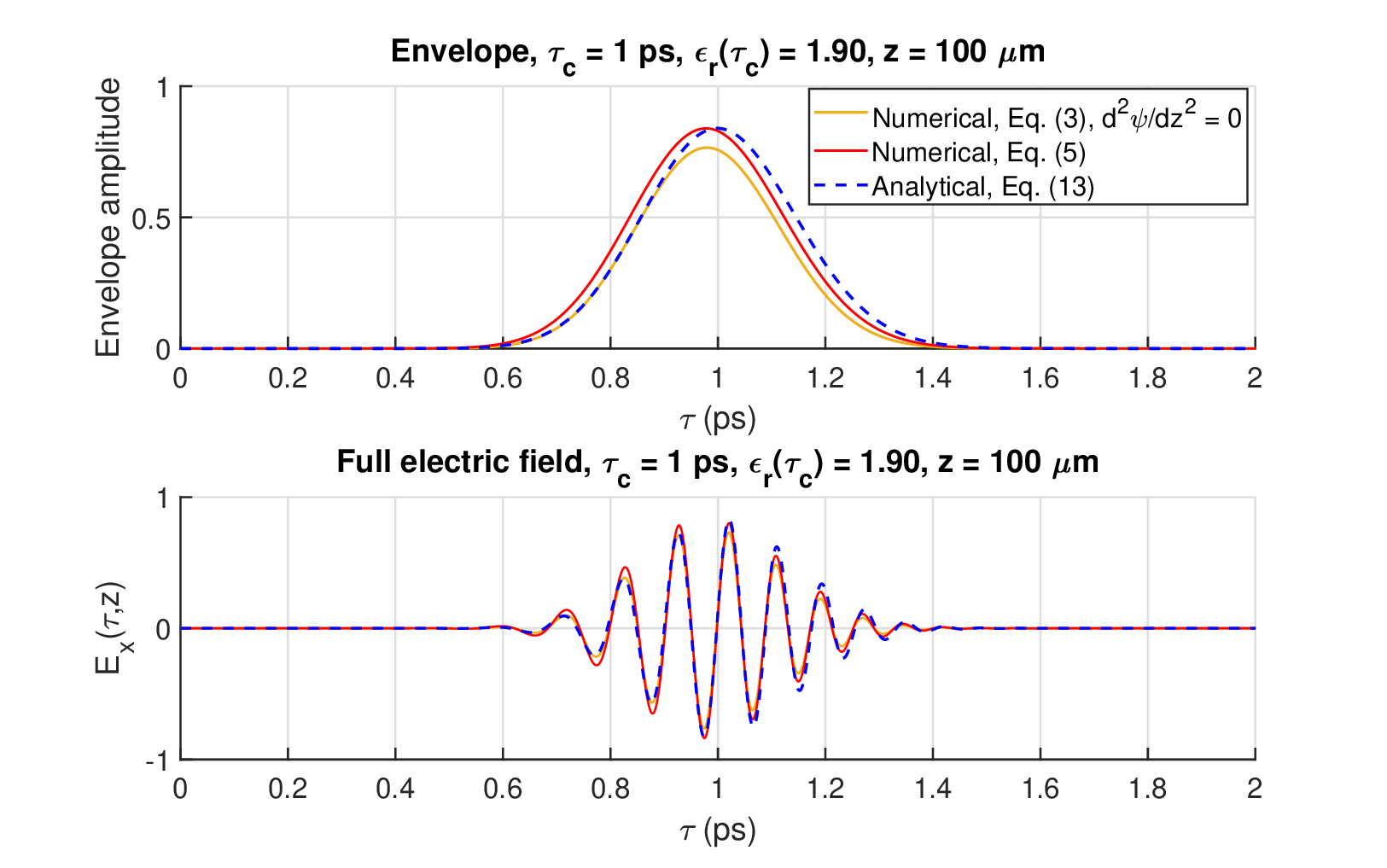}
    \caption{Gaussian pulse at $z = 100\,\mu$m for $\tau_{\text{c}} = 1\,\mu$s. Dashed blue line is the result provided by the analytical expression where $a(\tau_{\text{c}}), b(\tau_{\text{c}})$ remain constant. Dashed red line represents the solution obtained after solving Eq.(5) of the main manuscript. Dashed yellow line represents the solution yielded by Eq.(3) of the main manuscript. The top plot exhibits the pulse envelope, while the bottom plot shows the whole pulse.}
    \label{fig1_rev1_q2}
\end{figure}

\noindent Before evaluating the accuracy of the approximated analytical expression in Eq.~(13), let us estimate the temporal interval over which the pulses in Fig.~2 of the main manuscript  propagate. Throughout the individual duration of each pulse, the permittivity is assumed to be constant and is evaluated at the center $\tau_c$: $\varepsilon_{\text{r}}(\tau_{\text{c}}) = 1.5 + 4\times10^{11} \tau_{\text{c}}$. This temporal interval can be estimated as a function of the propagation distance $z$:
\begin{equation}\label{Dtau}
\Delta \tau (\tau_{\text{c}}) = \frac{z}{c/\sqrt{\varepsilon(\tau_{\text{c}})}}
\end{equation}
where $z = 100\,\mu\text{m}$ in our case. For the first pulse in Fig. 2 (blue-colored), where $\tau_{\text{c}} = 1\,\text{ps}$, we find $\Delta \tau(1\,\text{ps}) = 0.45\,\text{ps}$. Conversely, for the last pulse (green-colored), where $\tau_{\text{c}} = 9\,\text{ps}$, we obtain $\Delta \tau (9 \,\text{ps}) = 0.75\,\text{ps}$. In all considered cases, the interval $\Delta \tau$ does not exceed $1\,\text{ps}$. Within this window, we can evaluate the actual variation of the permittivity. Utilizing $\varepsilon_{\text{r}}(\tau) = 1.5 + 4\times 10^{11} \tau$, we obtain:
\begin{align}\label{De1ps}
\Delta \varepsilon_{\text{r}} (\tau_{\text{c}} = 1 ,\text{ps}) &= 0.18, \\
\label{De9ps} \Delta \varepsilon_{\text{r}} (\tau_{\text{c}} = 9 ,\text{ps}) &= 0.3,\end{align}
which leads to the following relative errors:
\begin{align}
\label{relative1} \frac{\Delta \varepsilon_{\text{r}}(\tau_{\text{c}} = 1, \text{ps})}{\varepsilon_{\text{r}}(\tau_{\text{c}} = 1, \text{ps})} &= 9.5\%, \\
\label{relative2} \frac{\Delta \varepsilon_{\text{r}}(\tau_{\text{c}} = 9, \text{ps})}{\varepsilon_{\text{r}}(\tau_{\text{c}} = 9, \text{ps})} &= 5.9\%\,.
\end{align}
These calculations demonstrate that the permittivity varies minimally within the interval under consideration, with the relative error never exceeding $10\%$. \\

\begin{figure}[!t]
    \centering  
    \subfigure[]{\includegraphics[width=0.7\columnwidth]{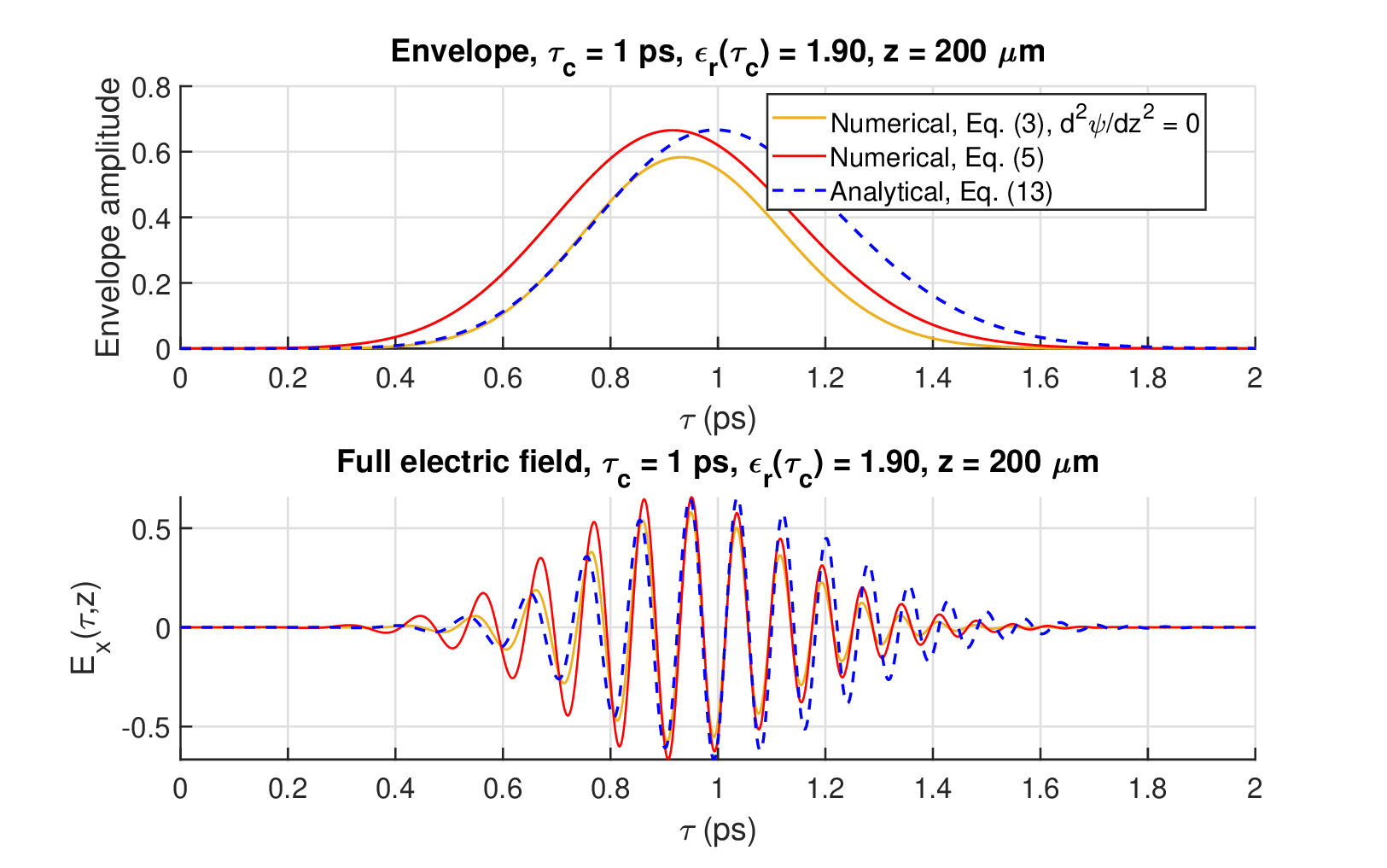}}
    \subfigure[]{\includegraphics[width=0.7\columnwidth]{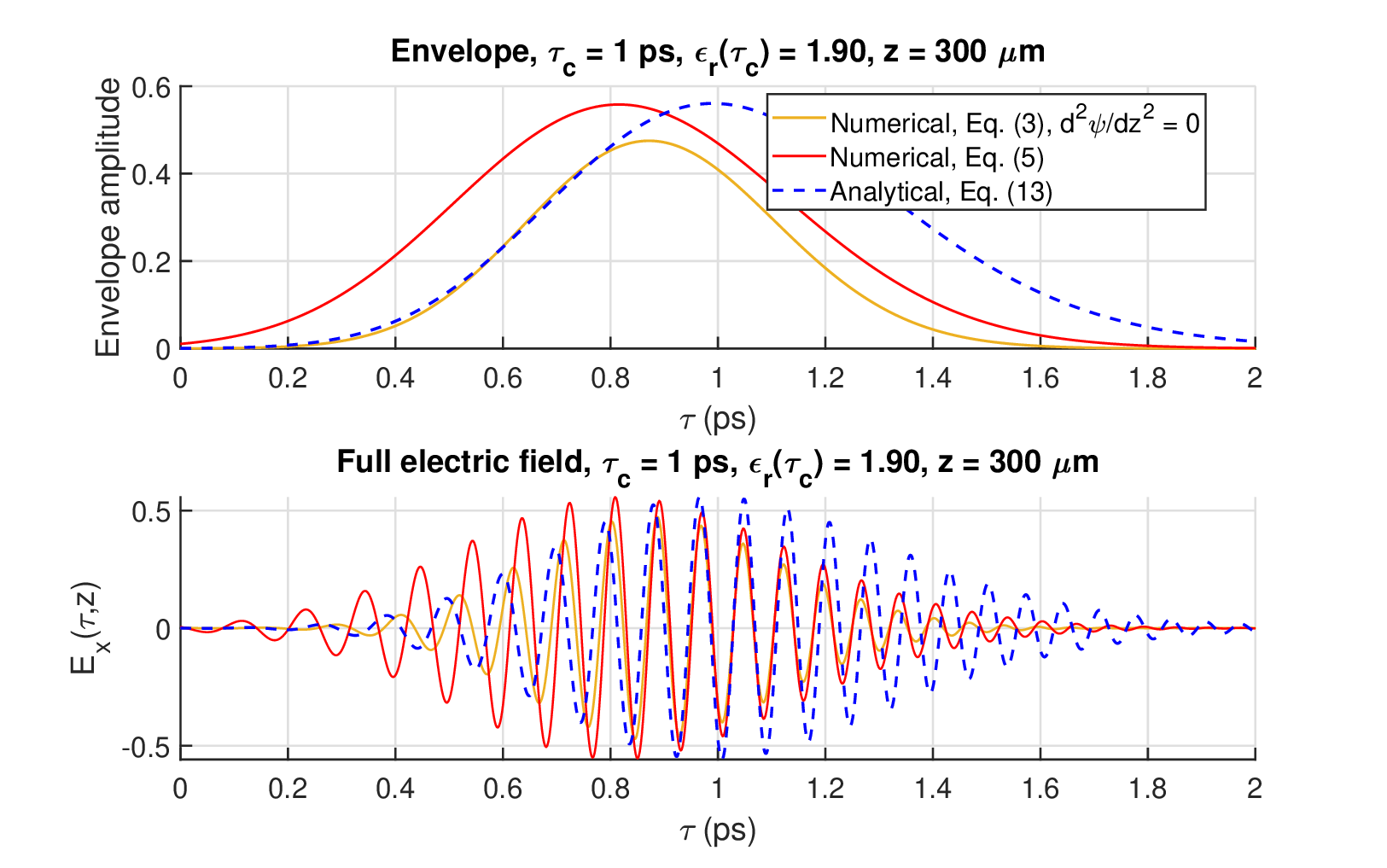}}
    \caption{ Representation of the analytical and numerical solutions for the case with $\tau_{\text{c}} = 1\,$ps for different distances $z$. From top to bottom: $z = 200\,\mu$m, $z = 300\,\mu$m.}
    \label{fig2_rev1_q2}
\end{figure}

\noindent With these figures in mind, we first focus on the scenario defined by $\tau_{\text{c}} = 1\,\text{ps}$, where $\varepsilon(\tau_{\text{c}}) = 1.9$, as shown in Fig.~\ref{fig1_rev1_q2}. Both the pulse envelope and the full waveform are represented. The analytical solution from Eq. (13) is compared to the numerical solutions obtained from Eq. (5) and Eq. (3)—the latter of which does not neglect time derivatives. The evaluation is carried out at the same propagation distance, $z = 100\,\mu\text{m}$, where the relative error remains below $10\%$, demonstrating good overall agreement. As expected, the analytical expression and the numerical solution of Eq. (5) are very similar, exhibiting only a slight shift in time. The most rigorous solution, derived from Eq. (3), deviates in a similar manner, but its amplitude is also affected, showing a slight reduction compared to the other cases. Furthermore, the envelope representation reveals a minor reduction in pulse width for this final case. Nevertheless, the general agreement remains quite strong.  \\

\noindent This small time shift, or temporal delay, may be primarily related to the pulse velocity. In the analytical model, the velocity is assumed to be constant, $v(\tau_{\text{c}}) = c/\sqrt{\varepsilon_{\text{r}} (\tau_{\text{c}})}$. Conversely, in the numerical solutions, the velocity is time-dependent—meaning $a(\tau)$ and $b(\tau)$ also vary with time—and the wave dynamics are governed by their respective wave equations. Overall, the numerical solutions appear to be slightly slower than the analytical one. However, for a propagation distance of $z = 100\,\mu\text{m}$, this difference remains negligible. \\

\noindent The error increases as the propagation distance $z$ increases too, as discussed in the main manuscript. Consequently, the assumption of a constant permittivity during pulse propagation loses validity at larger distances due to the growth of $\Delta \tau (\tau_{\text{c}})$. Notice that:
\begin{align}
\label{200} \Delta \varepsilon_{\text{r}} (z = 200\,\mu\text{m}) &= 0.36 \implies \frac{\Delta \varepsilon_{\text{r}}(\tau_{\text{c}} = 1\,\text{ps})}{\varepsilon_{\text{r}}(\tau_{\text{c}} = 1\,\text{ps})}  \approx 20\%, \\
\label{300}\Delta \varepsilon_{\text{r}} (z = 300\,\mu\text{m}) &= 0.55 \implies \frac{\Delta \varepsilon_{\text{r}}(\tau_{\text{c}} = 1\,\text{ps})}{\varepsilon_{\text{r}}(\tau_{\text{c}} = 1\,\text{ps})} \approx 30\%,
\end{align}
representing a variation of $20\%$ and $30\%$ in $\varepsilon(\tau)$, respectively. Figs.~\ref{fig2_rev1_q2}(a)-(b) display the pulses at $z = 200\,\mu\text{m}$ and $z = 300 \,\mu\text{m}$, respectively. The temporal shift becomes significant, indicating an increased delay between the numerical solutions and the analytical one. The delays of the two numerical solutions are no longer equal; the solution from (5) exhibits the largest delay. The additional terms in (3) seem to reposition the pulse between the analytical solution and the numerical solution from (5). Aside from this delay, the amplitude of the numerical solution from Eq.~(5) remains similar to that of the analytical function. 
To provide a fair comparison, let us estimate the delay for each case. It has been estimated by comparing the temporal position of the pulse peaks from the analytical and the solution from (5), which is the most delayed. These data, including the relative errors, are summarized in Table~\ref{tab1}:
\begin{table}[h]
\centering
\begin{tabular}{|c|c|c|c|}
\hline
$ z$ & $\tau_{\text{c}}$ & delay & relative error \\ \hline
$100\,\mu$m    & $1\,$ps  & $0.03\,$ps &  $3\%$\\ \hline
$200\,\mu$m    & $1\,$ps  & $0.1\,$ps  & $10\%$ \\ \hline
$300\,\mu$m   & $1\,$ps   & $0.2\,$ps  & $20\%$ \\ \hline
\end{tabular} \vspace{0.1cm}
\caption{Estimation of the relative error between analytical and numerical pulses for $\tau_{\text{c}} = 1\,$ps. }
\label{tab1}
\end{table}

\begin{figure}[!t]
    \centering  
    \vspace{-0.2cm}
    \subfigure[]{\includegraphics[width=0.6\columnwidth]{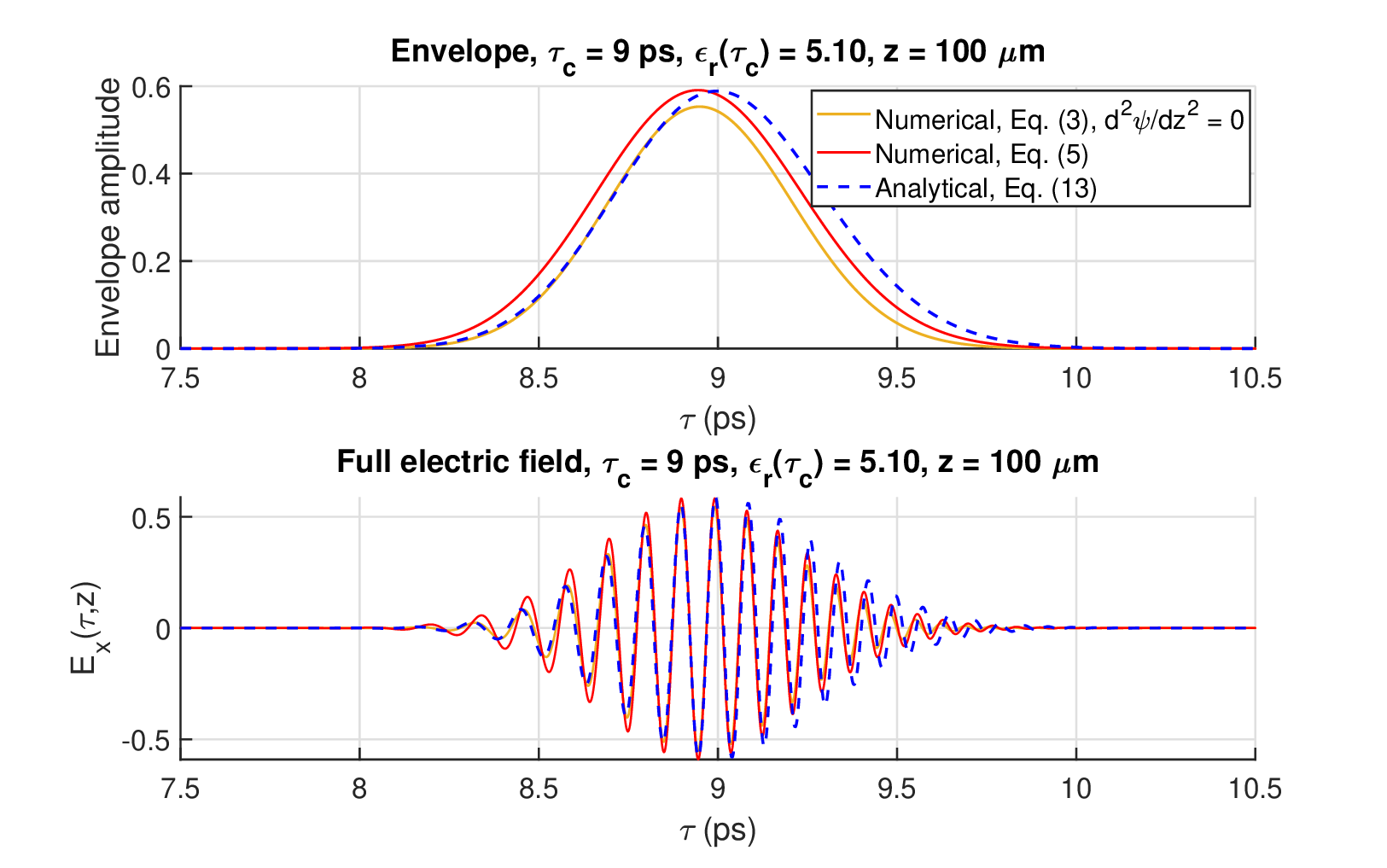}}
    \vspace{-0.2cm}
    \subfigure[]{\includegraphics[width=0.6\columnwidth]{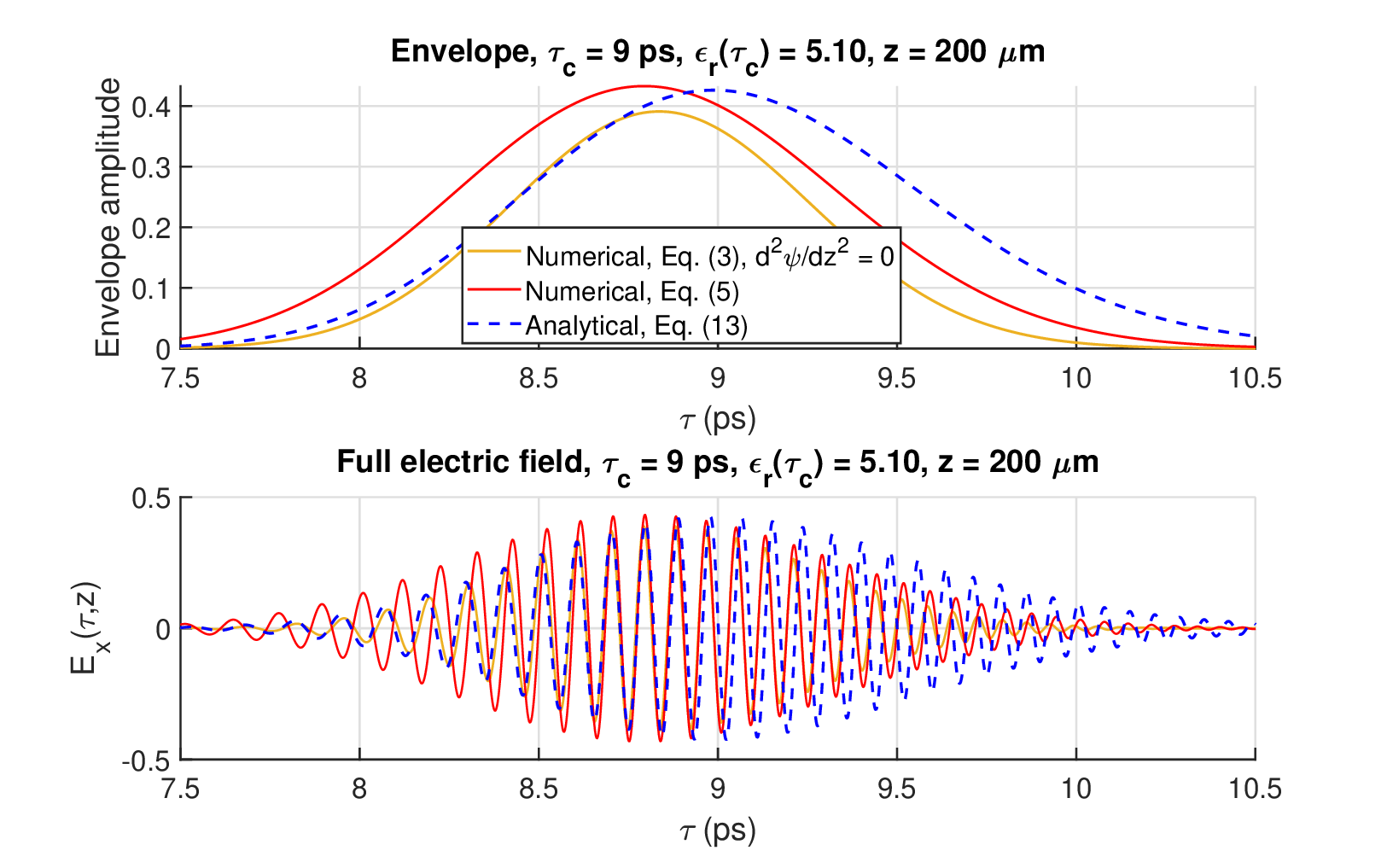}}
    \vspace{-0.2cm}
    \subfigure[]{\includegraphics[width=0.6\columnwidth]{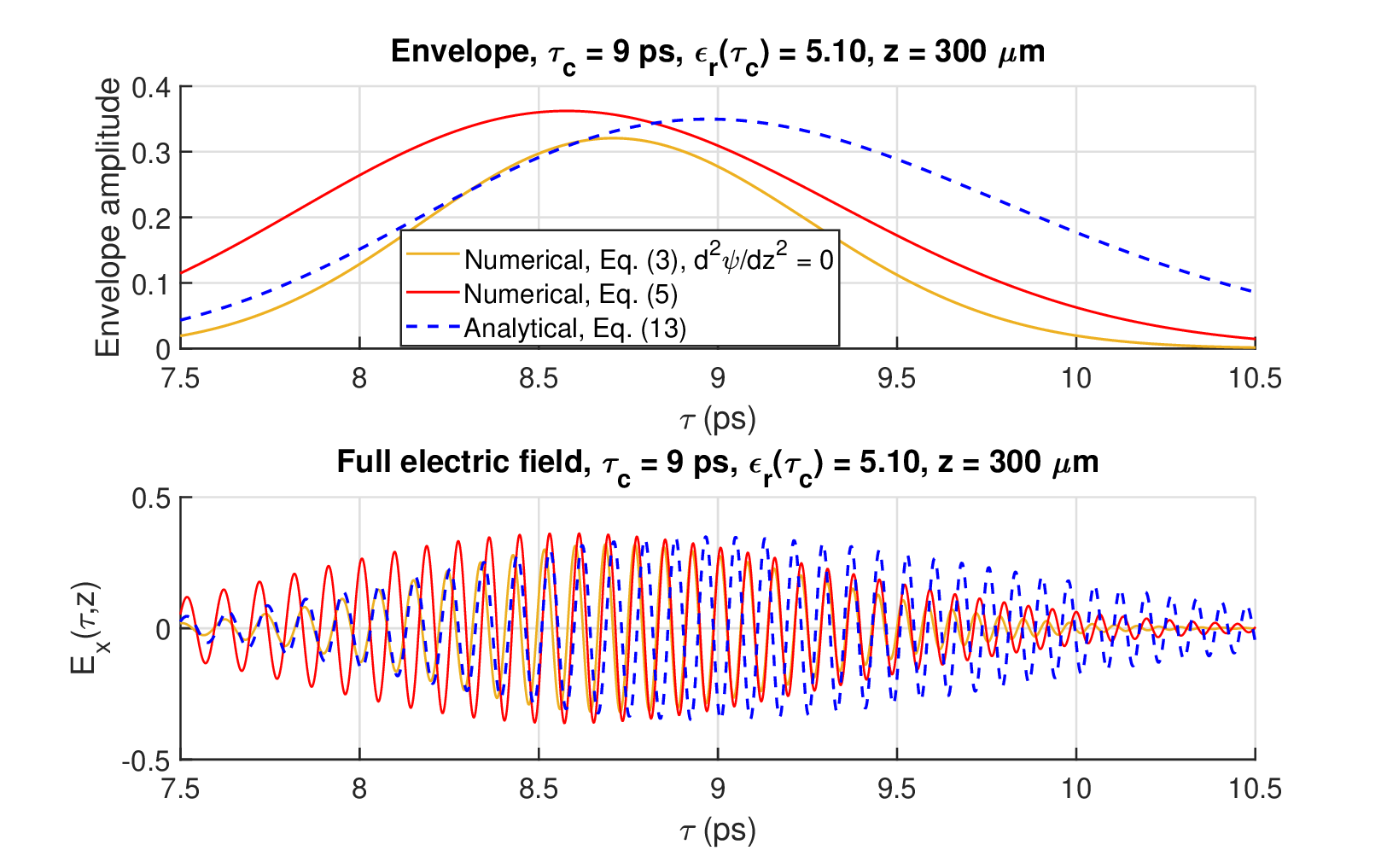}}
    \caption{Representation of the analytical and numerical solutions for the case with $\tau_{\text{c}} = 9\,$ps. Now, $\varepsilon_{\text{r}}(\tau_{\text{c}}) = 5.1$. From top to bottom: $z = 100\,\mu$m, $z = 200\,\mu$m, $z = 300\,\mu$m. }
    \label{fig3_rev1_q2}
\end{figure}

\noindent It is also important to mention that the solution from Eq. (3) also manifests a decrease of the pulse width, which is basically due to the more complicated dynamics of the wave equation. Now, time derivatives of the permittivity are no longer neglected and introduce additional terms to the equation. Their immediate effects are a slower width growth, and a reduction of the delay. Table~\ref{tab2} shows the percentage of width difference between the analytical solution and the solution from Eq. (3). The width estimation is carried out by assuming $\text{width} \approx 2\tau(A_{\text{max}}/2)$, with $A_{\text{max}}$ being the maximum amplitude of the envelope. 

\begin{table}[h]
\centering
\begin{tabular}{|c|c|c|c|c|}
\hline
$ z$ & $\tau_{\text{c}}$ & Analytical pulse width  & Pulse width from (3) & relative error \\ \hline
$100\,\mu$m    & $1\,$ps  & $0.315\,$ps &  $0.314\,$ps & 3\% \\ \hline
$200\,\mu$m    & $1\,$ps  & $0.523\,$ps  & $0.438\,$ps & 16.25\% \\ \hline
$300\,\mu$m   & $1\,$ps   & $0.727\,$ps  & $0.548\,$ps & 24.6\% \\ \hline
\end{tabular} \vspace{0.1cm}
\caption{Estimation of the relative error between analytical and numerical pulse from Eq. (3) for $\tau_{\text{c}} = 1\,$ps. }
\label{tab2}
\end{table}

\noindent Concerning the opposite extreme case with $\tau_{\text{c}} = 9\,$ps, the relative error of the permittivity at a distance of $z = 100\,\mu\text{m}$ is considerably smaller, as corroborated by the calculation in \eqref{relative2}. This finding is particularly interesting because it implies that the analytical approximation may remain valid over longer propagation distances. Figs.~\ref{fig3_rev1_q2} shows the three solutions at three different distances: $z = 100\,\mu\text{m}$, $z = 200\,\mu\text{m}$, and $z = 300\,\mu\text{m}$. Although the differences between the numerical and analytical solutions may appear significant at first glance, the evaluation of the relative error demonstrates that this case is remarkably more accurate than the previous one. As shown in Table~\ref{tab3}, the relative error barely exceeds $5\%$ in the worst-case scenario (at $z = 300\,\mu\text{m}$), whereas the same case yields a relative error of up to $20\%$ for $\tau_{\text{c}} = 1\,$ps. \\

\begin{table}[h]
\centering
\begin{tabular}{|c|c|c|c|}
\hline
$z$ & $\tau_{\text{c}}$ & delay & relative error \\ \hline
$100\,\mu$m    & $9\,$ps  & $0.05\,$ps &  $0.56\%$\\ \hline
$200\,\mu$m    & $9\,$ps  & $0.22\,$ps  & $2.4\%$ \\ \hline
$300\,\mu$m   & $9\,$ps   & $0.5\,$ps  & $5.6\%$ \\ \hline
\end{tabular} \vspace{0.1cm}
\caption{Estimation of the relative error between analytical and numerical pulses for $\tau_{\text{c}} = 9\,$ps.}
\label{tab3}
\end{table}

\noindent Concerning the pulse width, otherwise, the accuracy level decreases a bit. Table~\ref{tab4} shows a similar comparison as that carried out in Table~\ref{tab2}. Although the difference is larger for small values of $z$, the relative error at $z = 300\,\mu$m tends to converge around the $25\%$.    \\ 
\begin{table}[h]
\centering
\begin{tabular}{|c|c|c|c|c|}
\hline
$ z$ & $\tau_{\text{c}}$ & Analytical pulse width  & Pulse width from (3) & relative error \\ \hline
$100\,\mu$m    & $9\,$ps  & $0.670\,$ps &  $0.603\,$ps & 10\% \\ \hline
$200\,\mu$m    & $9\,$ps  & $1.23\,$ps  & $1\,$ps & 20\% \\ \hline
$300\,\mu$m   & $9\,$ps   & $1.7\,$ps  & $1.3\,$ps & 27\% \\ \hline
\end{tabular} \vspace{0.1cm}
\caption{Estimation of the relative error between analytical and numerical pulse from Eq. (3) for $\tau_{\text{c}} = 1\,$ps. }
\label{tab4}
\end{table}

\noindent A final aspect to discuss concerns the pulse amplitude. According to the results in both cases, the solutions from the differential equation in Eq.(3) exhibit an amplitude reduction compared to the others. At first glance, the effect of considering the temporal derivatives of $\varepsilon_{\text{r}}$ manifests as a decrease in the peak amplitude of the pulse. Table~\ref{tab5} summarizes the amplitude ratios, calculated by dividing the exact numerical amplitude by the analytical one. As can be observed, this ratio does not depend on the permittivity value itself, since the reduction is remarkably similar for the same propagation distance $z$. Given that this effect emerges directly when taking into account the temporal derivatives of $\varepsilon_{\text{r}}(\tau)$, it is straightforward to infer that the amplitude decay is primarily driven by the rate of change of the permittivity, which is identical in both cases. And that can be the reason why the amplitude ratio is the same for the same distance $z$. \\

\begin{table}[h]
\centering
\begin{tabular}{|c|c|c|}
\hline
$ z$ & $\tau_{\text{c}}$ & Amplitude ratio  \\ \hline
$100\,\mu$m    & $1\,$ps  & $0.9$ \\ \hline
$200\,\mu$m    & $1\,$ps  & $0.84\,$  \\ \hline
$300\,\mu$m   & $1\,$ps   & $0.75\,$   \\ \hline
$ z$ & $\tau_{\text{c}}$ & Amplitude ratio  \\ \hline
$100\,\mu$m    & $9\,$ps  & $0.9\,$ \\ \hline
$200\,\mu$m    & $9\,$ps  & $0.833\,$   \\ \hline
$300\,\mu$m   & $9\,$ps   & $0.73\,$   \\ \hline
\end{tabular} \vspace{0.1cm}
\caption{Estimation of the relative error between the amplitude of the analytical result and the numerical solution from Eq.(3) of the main manuscript}
\label{tab5}
\end{table}

\begin{figure}[!t]
\centering  
\includegraphics[width=0.7\columnwidth]{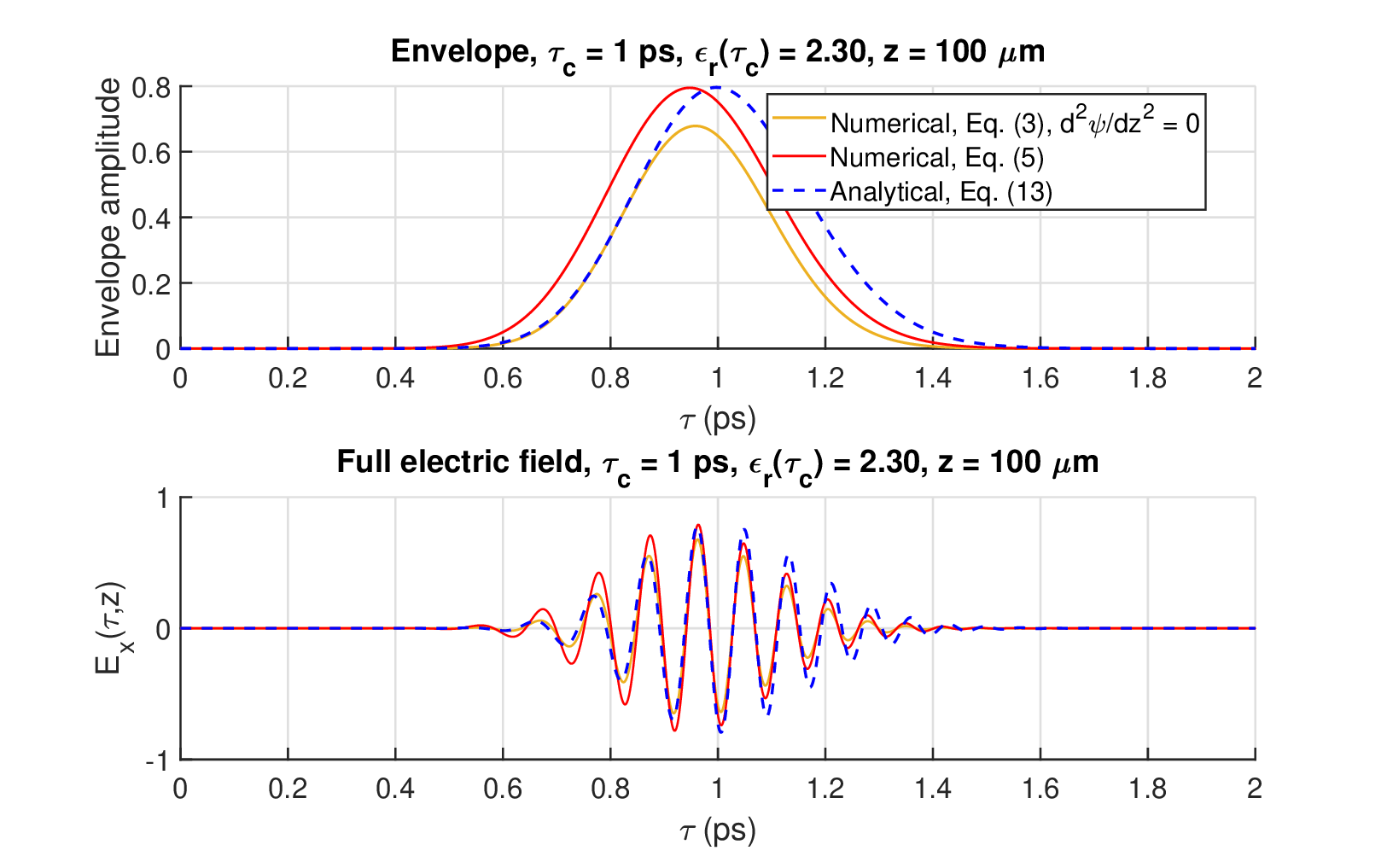}
\caption{Comparison between the three solutions when the slope increases up to $\alpha = 8\cdot 10^{11}\,$s. Now, for $\tau_{\text{c}} = 1\,\text{ps}$, $\varepsilon_{\text{r}}(\tau_{\text{c}}) = 2.3$. }
\label{fig4_rev1_q2}
\end{figure}

\noindent Finally, it is important to mention that the use of a linearly-dependent permittivity allows us to extend the discussion to the modulation ratio, beyond the propagation distance $z$. According to the linear expression of the permittivity, the modulation ratio is proportional to the slope of the expression. In all the cases analyzed above and in the main manuscript, the slope is $\alpha = 4\cdot 10^{11}\,\text{s}^{-1}$. The permittivity changes during a time lapse $\Delta \tau$ can be calculated as $\Delta \varepsilon_{\text{r}} = \alpha \Delta \tau$. This is exactly what we have done in \eqref{De1ps} and \eqref{De9ps}, where $\Delta \tau$ has previously estimated as a function of $z$ in \eqref{Dtau}. This leads us to conclude that $\Delta \varepsilon_{\text{r}}$ is proportional to $\Delta \tau$, in a same way as $\Delta \varepsilon_{\text{r}}$ is proportional to $z$. \\ 

\noindent To evaluate this conclusion, let us double the slope $\alpha$ up to $\alpha = 8\cdot 10^{11}\,$s$^{-1}$,
\begin{equation}
\varepsilon_{\text{r}}(\tau) = 1.5 + 8\cdot 10^{11} \tau\,.
\end{equation}
Focusing on $\tau_{\text{c}} = 1\,$ps, and for $z = 100\,\mu$m, 
\begin{equation}
\Delta \tau (\tau_{\text{c}}) \approx 0.46\, \text{ps}
\end{equation}
thus
\begin{equation}
\Delta \varepsilon_{\text{r}} \approx 0.36
\end{equation}
what leads to a relative error
\begin{equation}
\Delta \varepsilon_{\text{r}}(\tau_{\text{c}}) / \varepsilon_{\text{r}}(\tau_{\text{c}}) \approx 20\%
\end{equation}
By inspecting \eqref{200}, this relative error coincides with that obtained for $z = 200\,\mu$m. In other words, the effect of doubling the distance is quite similar to the effect of doubling the modulation of the material. Thus, the same conclusions can apply to both cases. Fig.~\ref{fig4_rev1_q2} shows the three pulses, with similar deviations as the case in the top plot in Fig.~\ref{fig2_rev1_q2}. \\

\noindent In summary, we have observed that the discrepancies between the numerical and analytical solutions increase over the propagation distance (or modulation ratio), as expected. Regarding the temporal delay, the relative error exhibits higher sensitivity when the dielectric constant is smaller. In contrast, the variations in amplitude appear to be less sensitive to the baseline permittivity value itself, being driven instead by its rate of change. We can conclude that for the case under consideration, the selected propagation distance of $z = 100\,\mu\text{m}$ is appropriate for the reliable application of the analytical Gaussian pulse expression.

\end{document}